\documentclass[%
 aip,
 jmp,%
 amsmath,amssymb,
preprintnumbers,
reprint,%
twocolumn,
]{revtex4-1}

\usepackage{graphicx}
\usepackage{dcolumn}
\usepackage{bm}

\usepackage{epstopdf}
\usepackage{oubraces}

\begin{document}

\preprint{submitted to Astrophys.\ J, NORDITA-2016-5}

\title[A new simple model for stellar cycle]{A new simple dynamo model for stellar activity cycle}

\author{N. Yokoi}
\altaffiliation[]{Guest researcher at the National Astronomical Observatory of Japan (NAOJ) and the Nordic Institute for Theoretical Physics (NORDITA)}
\email{nobyokoi@iis.u-tokyo.ac.jp}
\affiliation{Institute of Industrial Science, University of Tokyo, Tokyo 153-8505, Japan}

\author{D. Schmitt}%
\altaffiliation[]{Deceased}
\affiliation{Max-Planck Institut f\"{u}r Sonnensystemforschung, G\"{o}ttingen 37077, Germany}%

\author{V. Pipin}%
\affiliation{Institute of Solar--Terrestrial Physics,Russian Academy of Science,  Irkutsk 664033, Russia}%

\author{F. Hamba}%
\affiliation{Institute of Industrial Science, University of Tokyo, Tokyo 153-8505, Japan}%

\date{31 March 2016}

\begin{abstract}
A new simple dynamo model for stellar activity cycle is proposed. By considering an inhomogeneous mean flow effect on turbulence, it is shown that turbulent cross helicity (velocity--magnetic-field correlation) should enter the expression of turbulent electromotive force as the coupling coefficient for the mean absolute vorticity. The inclusion of the cross-helicity effect makes the present model different from the current $\alpha$--$\Omega$-type models mainly in two points. First, in addition to the usual $\alpha$ (helicity effect) and $\beta$ (turbulent magnetic diffusivity), we consider the $\gamma$ coefficient (cross-helicity effect) as a key ingredient of the dynamo process. Second, unlike the $\alpha$ and $\beta$ coefficients, which are often treated as an adjustable parameter in the current studies, the spatiotemporal evolution of $\gamma$ coefficient should be solved simultaneously with the mean magnetic-field equations. The basic scenario for the stellar activity cycle in the present model is as follows: In the presence of turbulent cross helicity, the toroidal field is induced by the toroidal rotation in mediation by the turbulent cross helicity. Then, as usual models, the $\alpha$ or helicity effect generates the poloidal field from the toroidal one. This poloidal field, induced by the $\alpha$ effect, produces a turbulent cross helicity whose sign is opposite to the original one (negative production of cross helicity). The cross helicity with the opposite sign starts producing a reversed toroidal field. Eigenvalue analyses of the simplest possible present model give a butterfly diagram, which confirms the above scenario as well as the equator-ward migrations, the phase relationship between the cross helicity and magnetic fields, etc. 
These results suggest that the oscillation of the turbulent cross helicity is a key for the activity cycle. The reversal of the turbulent cross helicity is not the result of the magnetic-field reversal, but the cause of the latter. This new model is expected to open up the possibility of the mean-field or turbulence closure dynamo approaches.

\end{abstract}

\maketitle

\section{Introduction\label{sec:1}}
	Stellar magnetic activity cycle, in particular the solar one, has been investigated in an elaborative manner with several types of dynamo model \citep{bra2005,cha2010,cha2014}. Depending on the ingredients of the models, such as the field-generation mechanism, location for the dynamo action to work, role of the large-scale flows, etc., the dynamo model may be classified into several categories or classes. The mean-field $\alpha$--$\Omega$ dynamos \citep{par1955,mof1978,par1979,kra1980}, the interface dynamos \citep{par1993,tob1996}, the flux-transport dynamos \citep{wan1991,cho1995,dur1995,cho2011}, the nonlinear dynamic dynamos \citep{bra1989,tob1997,kle2000,kle2006,pip1999,pip2004,pip2011a,arl1999,wei2016}, are representative ones. 
	
	Among others, the migrating mean-field dynamo mechanism \citep{par1955,par1979} has been paid considerable attentions because of the novelty of its physical insights and the simplicity of the model structure. In the presence of helical properties of fluid motions, magnetic fields moving in the fluid can be twisted, leading to a possibility of the magnetic-field configuration that is perpendicular to the original magnetic-field one \citep{par1955}. This effect is usually called the $\alpha$ effect. Under this effect, the poloidal field can be generated from the twisted toroidal magnetic field. In addition to the $\alpha$ effect, if fluid motions perpendicular to the magnetic field are nonuniform with respect to the magnetic-field direction, the magnetic field advected by the inhomogeneous flow leads to a magnetic-field configuration that is perpendicular to the original one. This differential-rotation effect is often called the $\Omega$ effect. If the toroidal flow in the solar convection zone is nonuniform along the latitudinal direction, a toroidal magnetic field can be produced from a poloidal one. Combination of the $\alpha$ and $\Omega$ effects may lead to an oscillational behavior of the toroidal and poloidal magnetic-field components. On the basis of this notion, \citet{par1955} constructed a system of equations for the poloidal and toroidal magnetic-field components represented by the toroidal component of the vector potential, $A^\phi (\equiv A)$, and the toroidal component of the magnetic field, $B^\phi (\equiv B)$. In this system of equations (Parker equations), the statistical properties of small-scale or turbulent motions are represented by two transport coefficients, $\alpha$ and $\beta$. 
	
	The expressions for $\alpha$ and $\beta$ are obtained with the aid of a magnetohydrodynamic (MHD) turbulence closure theory. The expressions of $\alpha$ and $\beta$ depend on the closure theory adopted, ranging from simplest parameter expressions with no closure to elaborated ones based on a sophisticated MHD theory and modeling. The coefficient $\alpha$ represents the helical properties of turbulence. On the other hand, $\beta$ represents the enhanced or effective magnetic diffusivity due to turbulence. This effect is often called the turbulent magnetic diffusivity or anomalous resistivity. Under the assumption that the magnetic Reynolds number is low ($Rm \ll 1$ where $Rm = UL/\eta$, $U$: characteristic velocity, $L$: characteristic length, and $\eta$: magnetic diffusivity), $\alpha$ can be expressed in terms of the turbulent kinetic helicity $\langle {{\bf{u}}' \cdot \mbox{\boldmath$\omega$}'} \rangle$ [${\bf{u}}'$: velocity fluctuation, $\mbox{\boldmath$\omega$}' (= \nabla \times {\bf{u}}')$: vorticity fluctuation, $\langle {\cdots} \rangle$: ensemble average]. However, it is well-known that in the high-$Rm$ case, which is much more relevant to astrophysical phenomena, $\alpha$ is expressed not only by the kinetic helicity but also with the correction due to the turbulent current helicity $\langle {{\bf{b}}' \cdot {\bf{j}}'} \rangle$ [${\bf{b}}'$: magnetic-field fluctuation, ${\bf{j}}' (= \nabla \times {\bf{b}}')$: electric-current density fluctuation] \citep{pou1976}. Since $\alpha$ and $\beta$ represent how much magnetic field is generated and destroyed by turbulence, respectively, the evaluations of $\alpha$ and $\beta$, as well as the spatiotemporal distribution and evolution of the differential rotation, are essential information for the evolution of the Parker's equations. For a review of MHD closure theories in dynamos and more elaborated analysis of the inhomogeneous MHD turbulence, see \citet{bra2005,yos1990,yok2013a}.
	
	Not a few studies have been done in this framework of the Parker equations. By assuming some flow configurations compatible with the observation then, \citet{ste1969,ysmr1975a,ysmr1975b} succeeded in numerically reproducing the latitudinal evolution of the toroidal magnetic field (butterfly diagram). In order to obtain the equatorward migration of the toroidal magnetic field, it is required that the angular velocity should decrease as the radius increases in the solar convection zone ($\partial \omega_{\rm{F}} / \partial r < 0$ for the positive $\alpha$ in the northern hemisphere, $\omega_{\rm{F}}$: angular velocity). However, the developments of helioseismology revealed that the real radial distribution of the angular velocity in the solar convective zone does not satisfy this requirement. In addition to this point, Parker himself argued the limitation of the $\alpha$ dynamo in the solar convection zone from the viewpoint of magnetic-field buoyancy \citep{par1975}. Once the magnitude of a magnetic field has been amplified to some level, the magnetic flux tube starts moving upward from deeper to shallower regions. Since this buoyancy convection timescale, even for a very weak magnetic flux tube, is estimated very short as compared with the $\alpha$-dynamo amplification timescale, it reaches the solar surface without being sufficiently amplified by the dynamo. This is called the magnetic buoyancy dilemma intrinsic to the magnetic flux tube.

	These difficulties have been considered to be obviated by assuming that the dynamo action is operated in a convectively stable thin layer between the radiative and convection zones \citep{par1975}. This region is associated with a large differential rotation and is called the {\it{tachocline}}. In the tachocline dynamo picture, the solar magnetic field is generated by a strong differential rotation there, and may have enough time to be amplified by the $\alpha$ dynamo before it rises up by the magnetic buoyancy \citep{spi1980,bal1982}.

	As mentioned above, depending on the location of the field generation, the dynamo models are divided into two types; (i) the flux-transport dynamo, which operates in the overshoot layer below the bottom of convection zone; and (ii) the mean-field distributed dynamo, which works in the bulk of the convection zone in particular at the subsurface shear layer.
 	
	On the surface of the Sun, large-scale meridional circulations with the speed of $\sim$ 15\ m\ s$^{-1}$ in the poleward direction have been observed. The combination of the meridional circulation and the $\alpha$--$\Omega$ mean-field model demonstrated its ability to reproduce the basic features of the solar magnetic activity cycle such as the equatorward migration of the sunspots, the cycle period, the phase difference between the poloidal and toroidal magnetic fields. In the combination with meridional circulations or convective-diffusive motions such as the Babcock--Leighton-type convection, which produces the east-west tilt of the sunspot pairs under the action of the Coriolis force on the magnetic flux tube during its rise through the convection zone, it is expected that the tachocline dynamo can reproduce several properties of solar activity cycle \citep{cho1995,dik2006}. This approach is called the flux-transport dynamo. In practical numerical simulations of the model, the large-scale fluid motion is fixed according to the helioseismology observation data. 

	It should be noted that the thin flux tube simulations by \citet{dsi1993,fan1993,cal1995}  showed that, with their initial and boundary conditions, the magnetic field energy is greater than the turbulent equipartition energy inside the convection zone, so it is hard for the toroidal field lines represented by the flux tube to be twisted by the helical turbulence. This suggests that the $\alpha$--$\Omega$ dynamo is unlikely to operate to the flux tube inside the solar convection zone. At the same time, there are some unresolved problems on the flux-transport model results from the active-region observations, which include that (i) emerging bipolar active regions have not tilt, which is developed after while \citep{ste2012a,ste2012b}; (ii) the evolution of tilt depend on the size of the active regions \citep{tla2013,mcc2016};  (iii) the active regions rotate slower than surrounding plasma with angular velocity that corresponds to the near-surface shear layer \citep{ben1999}; (iv) the most effect to the relative sunspot number comes from small active regions \citep{ste2012b}; (v) the long-term evolution of tilt in the solar cycle is still open \citep{das2010,tla2013}. 

	As for the meridional circulation, however, its spatial pattern is still observationally open. It has been reported that the behavior of dynamo magnetic field highly depends on the flow pattern of the meridional circulation adopted in the numerical simulation \citep{bon2006,jou2007}. The current observational status are seen in \citet{zha2013,sch2013,raj2015}. The flux transport dynamo with very complicated meridional flow patterns was investigated, and it was shown that if an equatorward flow in low latitudes at the bottom of the convective zone is present, the flux transport dynamo works even in the case of a meridional circulation with very complicated flow structures \citep{haz2014}. Similar works are seen for dynamos with double and multiple cells \citep{pip2013,pip2014a}, and dynamos in the global 3D simulations \citep{rac2011,war2014}.
	
	Apart from the practical predictability of the solar-activity cycle, there are still open issues in the flux-transport dynamo models. Some of them are related to the evaluation of spatiotemporal distributions of turbulence, its effects and boundary conditions in the solar convection zone. For example, one of the essential ingredients of the flux-transport dynamo in general is the requirement for the spatial distribution of the effective or turbulent magnetic diffusivity: the turbulent magnetic diffusivity should decrease as the radial position becomes deeper. Because of this requirement, the magnetic field moves in the equatorward direction reflecting the large-scale circulation motion in the deeper region. Otherwise, the magnetic field may move in the poleward direction reflecting the circulation motion in the shallow region.
	
	As has been seen in the above descriptions on the radial dependence of the turbulent magnetic diffusivity $\beta$, basic behaviors of the large-scale magnetic field highly depend on the turbulent transport coefficients $\alpha$ and $\beta$, whose spatiotemporal evolutions are still not fully understood. In the current flux-transport dynamo model, the spatial distribution of the turbulent magnetic diffusivity $\beta$ is prescribed, and the turbulent helicity-related transport coefficient $\alpha$ is often treated as an adjustable parameter. An interesting point is the role of the magnetic fluctuations in magnetic-field reversal \citep{cho2007,cho2011}. By comparing the observation of the photospheric magnetic-field data with the predictions from a simple dynamo model with magnetic fluctuations, it was shown that the magnetic fluctuation level is highly connected with the behavior of the solar dipole during the polarity reversal \citep{pip2014b}. 
	
	One possible alternative approach is to consider the spatiotemporal distributions of $\alpha$ and/or $\beta$ by solving the transport equations of them. Since the evolutions of $\alpha$, $\beta$, etc.\ depend on the mean fields, ${\bf{B}}$, $\nabla \times {\bf{B}}$, etc., the inclusion of the transport equations of them enables the incorporation of the nonlinear dynamics of dynamo into the model. In this line of thought, \citet{sch1995} introduced the dynamic part of the $\alpha$ coefficient, $\alpha^\ast$, in addition to the given kinetic part $\alpha_{\rm{L}}$, and constructed a simple transport equation of $\alpha^\ast$. By solving the equation of $\alpha^\ast$ as well as the equations for the toroidal components of the vector potential and magnetic field, $A$ and $B$, the phase difference between the poloidal and toroidal magnetic fields was reproduced in the numerical simulation. Related to the inviscid invariance of the total magnetic helicity $\int_V {{\bf{a}} \cdot {\bf{b}}} dV$, it has some meaning to construct a transport equation of the current helicity density $\langle {{\bf{b}' \cdot {\bf{j}}'}} \rangle$ part of $\alpha$, $\alpha_{\rm{m}}$. \citet{kle1995} adopted a simple model equation for $\alpha_{\rm{m}}$ and estimated the magnitude of the dynamo-generated magnetic field in the solar-type convection zones.
	
	In this paper, we suggest a new dynamo mechanism without showing any preference to the flux-transport dynamo scenario itself. We shall focus our attention on the inhomogeneous large-scale flow effect on the turbulent electromotive force (EMF). As we see in the following section, such an effect in the turbulent EMF is represented by the turbulent cross helicity (cross-correlation between the velocity and magnetic fluctuations) coupled with the mean or large-scale vorticity. Unlike the helicity and energy-related transport coefficients, $\alpha$ and $\beta$, the cross-helicity-related transport coefficient, $\gamma$, does change its sign during the magnetic-field reversal. In this sense, the spatiotemporal behaviors of the turbulent cross helicity is directly connected to the evolution of magnetic fields including the polarity reversal. 
	
	The cross helicity itself has been investigated in several contexts in space and astrophysical turbulence \citep{yok2013a}. Plasma relaxation with a cross-helicity constraint \citep{bis1993}, the large-scale behavior of cross helicity in the solar-wind turbulence \citep{yok2008,yok2011b}, 
and the trasnport enhancement and suppression in the turbulent magnetic reconnection \citep{yok2011c,hig2013,yok2013b,wid2016} are representative subjects of such investigations. In the context of the solar dynamo and activity cycle, some studies have been done related to the cross helicity. On the basis of an expression for the turbulent cross-helicity and/or the evolution equation of the turbulent cross helicity, the dynamic evolution of the turbulent cross helicity under the effects of the density stratification, large-scale magnetic fields, differential rotation, etc.\ were examined in the frameworks of a mean-field dynamo and of a flux-transport dynamo \citep{kle2003,kuz2007,pip2011b}. An attempt to measure the cross helicity on the surface of the Sun by observation has been also reported \citep{zha2011}.
	
	A dynamo model, where the poloidal and toroidal magnetic-fields are induced by the helicity and cross-helicity effects, respectively, had been applied to the solar magnetic-field reversal problem \citep{yos2000}. Although the dynamics of the toroidal-field evolution was not solved there, an oscillational behavior of the magnetic field was predicted with the aid of a very simple toy model. With this point in mind, in addition to the mean magnetic-field equations, in this work we shall  include the transport equation of the turbulent cross helicity rather than the counterparts of the turbulent energy and/or helicity.

	The organization of this paper is as follows. In Section~\ref{sec:2}, we point out the importance of the mean-flow inhomogeneity in treating turbulence, and how this inhomogeneity leads to the cross-helicity effect. In Section~\ref{sec:3}, the dynamo equations with the cross-helicity effect are presented and reduced to the simplest form with several assumptions and boundary conditions. In Section~\ref{sec:4}, an eigenvalue analysis is carried out and the critical condition for the oscillation of the magnetic field and the turbulent cross helicity is argued. In Section~\ref{sec:5}, the butterfly diagram of the magnetic field is discussed with special reference to the cross-helicity effect. The concluding remarks are given in Section~\ref{sec:6}.




\section{Flow inhomogeneity and cross-helicity effect\label{sec:2}}
	The inhomogeneities of the mean or large-scale fields are important ingredients providing a free-energy source for turbulence. For example, the mean-velocity shear, or strictly speaking, the mean-velocity strain, is considered to be one of the key ingredients producing turbulent energy. In the absence of the large-scale flow shear, we need some external source or forcing to generate and sustain turbulence. Here, we focus on the mean-flow inhomogeneity in dynamo and turbulence, and examine the resultant effect on the turbulent electromotive force (EMF) in the global magnetic-field equation.

\subsection{Mean flow inhomogeneity in dynamos}
	The equation of the mean magnetic field ${\bf{B}}$ is written as
\begin{equation}
	\frac{\partial {\bf{B}}}{\partial t}
	= \nabla \times \left( {{\bf{U}} \times {\bf{B}}} \right)
	+ \nabla \times {\bf{E}}_{\rm{M}}
	+ \eta \nabla^2 {\bf{B}},
	\label{eq:mean_mag_eq}
\end{equation}
where ${\bf{U}}$ is the mean velocity, $\eta$ is the molecular magnetic diffusivity, and ${\bf{E}}_{\rm{M}}$ is the turbulent electromotive force (EMF) defined by
\begin{equation}
	{\bf{E}}_{\rm{M}}
	\equiv \langle {{\bf{u}}' \times {\bf{b}}'} \rangle
	\label{eq:emf_def}
\end{equation}
(${\bf{u}}'$: velocity fluctuation, ${\bf{b}}'$: magnetic fluctuation, $\langle \cdots \rangle$: ensemble average), which represents the turbulence effect in the mean magnetic-field induction. In the first term of Eq.~(\ref{eq:mean_mag_eq}), the motions parallel to the mean magnetic field has no contributions, so we rewrite the first term on the right-hand side (r.h.s.) as
\begin{eqnarray}
	\lefteqn{
	\nabla \times \left( {{\bf{U}} \times {\bf{B}}} \right)
	= \nabla \times \left( {{\bf{U}}_\perp \times {\bf{B}}} \right)
	}\nonumber\\
	&= - \left( {{\bf{U}}_\perp \cdot \nabla} \right) {\bf{B}}
	- {\bf{B}} \left( {\nabla \cdot {\bf{U}}_\perp} \right)
	+ \left( {{\bf{B}} \cdot \nabla} \right) {\bf{U}}_\perp,
	\label{eq:mag_adv}
\end{eqnarray}
where ${\bf{U}}_\perp$ is the mean velocity perpendicular to the mean magnetic field. The first term in Eq.~(\ref{eq:mag_adv}) represents the advective derivative of the mean magnetic field due to the mean velocity. The second term represents the mean magnetic-field reduction (or induction) due to the divergence ($\nabla \cdot {\bf{U}}_\perp > 0$) [or convergence ($\nabla \cdot {\bf{U}}_\perp < 0$)] of the mean velocity. The third term indicates that the mean magnetic field is induced by a mean flow in the direction of the mean velocity if the velocity is inhomogeneous along the mean magnetic field (the differential-rotation effect). In the dynamo studies, this differential-rotation effect, often called the $\Omega$ effect, has been considered to play an essential role in producing the toroidal component of the mean magnetic field from the poloidal one.

	Next we consider the evolution of the fluctuation fields. The equations of the velocity and magnetic fluctuations, ${\bf{u}}'$ and ${\bf{b}}'$, are given as
\begin{eqnarray}
	\lefteqn{
	\frac{D{\bf{u}}'}{Dt}
	\left[ { \equiv \left( {
		\frac{\partial}{\partial t}
		+ {\bf{U}} \cdot \nabla} \right) {\bf{u}}'
	} \right]
	}\nonumber\\
	&&\hspace{10pt} = ({\bf{B}} \cdot \nabla) {\bf{b}}'
	+ ({\bf{b}}' \cdot \nabla) {\bf{B}}
	- ({\bf{u}}' \cdot \nabla) {\bf{U}}
	+ \cdots,
	\label{eq:fluct_u_eq}
\end{eqnarray}
\begin{eqnarray}
	\lefteqn{
	\frac{D{\bf{b}}'}{Dt}
	\left[ { \equiv \left({
	\frac{\partial}{\partial t}
		+ {\bf{U}} \cdot \nabla
	} \right) {\bf{b}}'
	} \right]
	}\nonumber\\
	&& \hspace{10pt} = ({\bf{B}} \cdot \nabla) {\bf{u}}'
	- ({\bf{u}}' \cdot \nabla) {\bf{B}}
	+ ({\bf{b}}' \cdot \nabla) {\bf{U}}
	+ \cdots.
	\label{eq:fluct_b_eq}
\end{eqnarray}
In the case that the mean velocity ${\bf{U}}$ is assumed to be uniform ${\bf{U}} = {\bf{U}}_0$ (${\bf{U}}_0$: uniform mean velocity), and the inhomogeneity of the mean velocity is entirely neglected, Eqs.~(\ref{eq:fluct_u_eq}) and (\ref{eq:fluct_b_eq}) are reduced to
\begin{equation}
	\frac{D{\bf{u}}'}{Dt}
	= ({\bf{B}} \cdot \nabla) {\bf{b}}'
	+ ({\bf{b}}' \cdot \nabla) {\bf{B}}
	+ \cdots,
	\label{eq:fluct_u_red_eq}
\end{equation}
\begin{equation}
	\frac{D{\bf{b}}'}{Dt}
	= ({\bf{B}} \cdot \nabla) {\bf{u}}'
	- ({\bf{u}}' \cdot \nabla) {\bf{B}}
	+ \cdots.
	\label{eq:fluct_b_red_eq}
\end{equation}

	The temporal evolution of the turbulent electromotive force ${\bf{E}}_{\rm{M}}$ [Eq.~(\ref{eq:emf_def})], in general, is given by the equations for the velocity and magnetic-field fluctuations as
\begin{equation}
	\frac{D}{Dt} \left\langle {{\bf{u}}' \times {\bf{b}}'} \right\rangle
	= \left\langle {
		\frac{D{\bf{u}}'}{Dt} \times {\bf{b}}'
		+ {\bf{u}}' \times \frac{D{\bf{b}}'}{Dt}
	} \right\rangle.
	\label{eq:emf_cal}
\end{equation}
If we adopt Eqs.~(\ref{eq:fluct_u_red_eq}) and (\ref{eq:fluct_b_red_eq}) for the fluctuation equations, ${\bf{E}}_{\rm{M}}$ has no direct dependence on the mean-velocity inhomogeneities. The often-adopted Ansatz: ``The turbulent electromotive force should be expressed in terms of the expansion of linear functions of the mean magnetic field'' as
\begin{equation}
	\langle {{\bf{u}}' \times {\bf{b}}'} \rangle^\alpha
	= \alpha^{\alpha a} B^a
	+ \beta^{\alpha ab} \frac{\partial B^a}{\partial x^b}
	+ \cdots,
	\label{eq:emf_ansatz}
\end{equation}
should be understood in this context of neglecting the mean-velocity inhomogeneities in turbulence equations.

	If we compare the fluctuation equations [Eqs.~(\ref{eq:fluct_u_red_eq}) and (\ref{eq:fluct_b_red_eq})] with the mean-field equation [Eq.~(\ref{eq:mean_mag_eq})], it is obvious that the treatments of the mean-velocity inhomogeneities in fluctuation and mean equations are remarkably different. In the mean magnetic-field evolution, an inhomogeneous mean flow is considered to play an essential role in magnetic field generation. On the other hand, in the fluctuation evolution, the effects of inhomogeneous mean flow are completely neglected.

\subsection{Cross-helicity effect}
\subsubsection{Mean flow inhomogeneity and cross helicity}
	The turbulent cross helicity is the cross-correlation between the velocity and magnetic-field fluctuations defined by
\begin{equation}
	W \equiv \langle {{\bf{u}}' \cdot {\bf{b}}'} \rangle.
	\label{eq:cross_helicity_def}
\end{equation}
If we retain the mean-velocity inhomogeneities in the fluctuation equations [Eqs.~(\ref{eq:fluct_u_eq}) and (\ref{eq:fluct_b_eq})], from Eq.~(\ref{eq:emf_cal}), we naturally have contributions of the cross helicity to the turbulent EMF as
\begin{eqnarray}
	\tau \lefteqn{
	\left\langle {
	{\bf{u}}' \times [({\bf{b}}' \cdot \nabla) {\bf{U}}]
	+ [({\bf{u}}' \cdot \nabla) {\bf{U}}] \times {\bf{b}}' 
	} \right\rangle^\alpha
	}\nonumber\\
	&& \hspace{20pt} = \epsilon^{\alpha ab} \tau \langle {u'{}^a b'{}^c} \rangle 
		\frac{\partial U^b}{\partial x^c}
	- \epsilon^{\alpha ba} \tau \langle {b'{}^a u'{}^c} \rangle 
		\frac{\partial U^b}{\partial x^c}
	\nonumber\\
	&& \hspace{20pt} = \tau \left( {
		\langle {u'{}^a b'{}^c} \rangle + \langle {u'{}^c b'{}^a} \rangle
	} \right) \epsilon^{\alpha ab} \frac{\partial U^b}{\partial x^c},
	\label{eq:crss_hel_effect_cal}
\end{eqnarray} 
where $\tau$ is the characteristic time scale of turbulence. If we adopt the isotropic representation of the cross-correlation tensor:
\begin{equation}
	\langle {u'{}^a b'{}^c} \rangle + \langle {u'{}^c b'{}^a} \rangle
	= \frac{2}{3} \delta^{ac} \langle {{\bf{u}}' \cdot {\bf{b}}'} \rangle
	\label{eq:crss_corl_iso}
\end{equation}
for simplicity, Eq.~(\ref{eq:crss_hel_effect_cal}) is reduced to
\begin{eqnarray}
	\tau \lefteqn{
	\left\langle {
	{\bf{u}}' \times [({\bf{b}}' \cdot \nabla) {\bf{U}}]
	+ [({\bf{u}}' \cdot \nabla) {\bf{U}}] \times {\bf{b}}' 
	} \right\rangle^\alpha
	}\nonumber\\
	&& \hspace{20pt} = \frac{2}{3} \tau \langle {{\bf{u}}' \cdot {\bf{b}}'} \rangle 
		(\nabla \times {\bf{U}})^\alpha.
	\label{eq:crss_hel_effect_cal_iso}
\end{eqnarray}
Of course, depending on the statistical properties of turbulence, the cross-correlation should have been expressed in a more elaborated form. In this sense, the isotropic representation [Eq.~(\ref{eq:crss_corl_iso})] is too much simplified. However, we still see that the primary quantity that is coupled with the mean-flow inhomogeneity is the cross helicity or the cross-correlation between the velocity and magnetic-field fluctuations. For the elaborated expression for the mean vorticity-related term in the turbulent EMF, see \cite{yos1990} and \cite{yok2013a}.

	The above argument clearly shows that, if we retain the mean-flow inhomogeneity in the turbulence equations, we should include the cross-helicity effect into the the expression for the turbulent EMF, in addition to the usual effects such as the turbulent magnetic diffusivity $\beta$ and the helicity or $\alpha$ effect. Then turbulent EMF should be expressed as
\begin{equation}
	{\bf{E}}_{\rm{M}} 
	= - \beta {\bf{J}} 
	+ \alpha {\bf{B}} 
	+ \gamma \mbox{\boldmath$\Omega$}.
	\label{eq:emf_model_full}
\end{equation}
Here, $\beta$ is the turbulent magnetic diffusivity, which is related to the turbulent MHD energy, $\alpha$ is the helicity or $\alpha$ effect, which is related to the turbulent residual helicity (the difference of the kinetic and current helicity), and $\gamma$ is the cross-helicity effect, which is related to the turbulent cross helicity. Simplified relationships are written as
\begin{subequations}\label{eq:abc_model}
\begin{equation}
	\beta 
	= C_\beta \tau \langle {{\bf{u}}'{}^2 + {\bf{b}}'{}^2} \rangle /2
	\equiv C_\beta \tau K,
	\label{eq:beta_model}
\end{equation}
\begin{equation}
	\alpha 
	= C_\alpha \tau \langle {
		- {\bf{u}}' \cdot \mbox{\boldmath$\omega$}'
		+ {\bf{b}}' \cdot {\bf{j}}'
		} \rangle
	\equiv C_\alpha \tau H,
	\label{eq:alpha_model}
\end{equation}
\begin{equation}
	\gamma 
	= C_\gamma \tau \langle {{\bf{u}}' \cdot {\bf{b}}'} \rangle
	\equiv C_\gamma \tau W,
	\label{eq:gamma_model}
\end{equation}
\end{subequations}
where $\tau$ is the characteristic time scale of turbulence and $C_\beta$, $C_\alpha$, and $C_\gamma$ are the model constants. More elaborated expressions for these transport coefficients and their relationship to Eq.~(\ref{eq:abc_model}) are given in Appendix~\ref{sec:a1}.

\subsubsection{Physical origins\label{sec:cr_phys}}
	Equation~(\ref{eq:emf_cal}) with Eq.~(\ref{eq:crss_hel_effect_cal}) or (\ref{eq:crss_hel_effect_cal_iso}) shows that the cross-correlations between the velocity and magnetic fluctuations coupled with the mean-flow inhomogeneities lead to the cross-helicity effect in the turbulent EMF. In order to get clear understanding of the cross-helicity effect, here we shall feature the physical processes and conditions that cause this effect.
	
\begin{figure}[htb]
\begin{center}
\includegraphics[width=.35\textwidth]{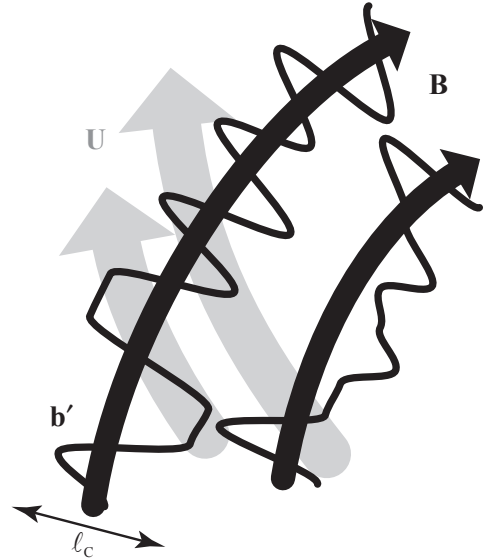}
\caption{\label{fig:random_fld} Schematic picture of the energy-containing scale $\ell_{\rm{C}}$ and inhomogeneous mean fields. ${\bf{B}}$: inhomogeneous mean magnetic field, ${\bf{b}}'$: magnetic fluctuation, ${\bf{U}}$: inhomogeneous mean flow. The fluctuating magnetic field ${\bf{b}} (= {\bf{B}} + {\bf{b}}')$ is depicted by distorted solid thin curves.}
\end{center}
\end{figure}

	As we saw the expressions in Eq.~(\ref{eq:abc_model}) [and its generalized ones in Eq.~(\ref{eq:abc_exp_wave})], the transport property of turbulence is mainly determined by the largest-scale fluctuations or the energy-containing eddies. Since our main interests lie in the turbulent transport represented by the turbulent EMF, we shall here suppose that the random fluctuations are characterized by the length scale $\ell_{\rm{C}}$ of the largest-scale fluctuations, which is roughly equal to the energy-containing eddies. The mean fields can be inhomogeneous within the dimension of $\ell_{\rm{C}}$ as schematically depicted in Fig.~\ref{fig:random_fld}.

	In hydrodynamic turbulence, it has been established that a non-uniform mean flow will affect the turbulent transport much. For instance, it is well-known that the inhomogeneous mean flow gives an essential contribution to the momentum transport through the Reynolds stress, which is often expressed in terms of the eddy-viscosity representation after Boussinesq as
\begin{equation}
	\langle {u'{}^\alpha u'{}^\beta} \rangle_{\rm{D}}
	= - \nu_{\rm{T}} \left( {
		\frac{\partial U^\alpha}{\partial x^\beta}
		+ \frac{\partial U^\beta}{\partial x^\alpha}
	} \right),
	\label{eq:Rey_strss_eddy_visc}
\end{equation}
where $\nu_{\rm{T}}$ is the eddy or turbulent viscosity and $\mbox{\boldmath$\cal{A}$}_{\rm{D}}$ denotes the traceless or deviatoric part of the tensor defined by ${\cal{A}}^{\alpha\beta}_{\rm{D}} = {\cal{A}}^{\alpha\beta} - A^{aa} \delta^{\alpha\beta} /3$. Note that Eq.~(\ref{eq:Rey_strss_eddy_visc}) is one of the simplest heuristic models for the Reynolds stress. As for the theoretical derivation of the Reynolds-stress expression [Eq.~(\ref{eq:Rey_strss_eddy_visc}) and further] with the analytical expressions of the transport coefficients, the reader is referred to \cite{yos1984} (for mirrorsymmetric case) and \cite{yok1993,yok2016} (for non-mirrorsymmetric case).

	In what follows, we focus our attention on the mean-flow inhomogeneity effect on the magnetic-field transport.

\begin{figure}[htb]
\begin{center}
\includegraphics[width=.40\textwidth]{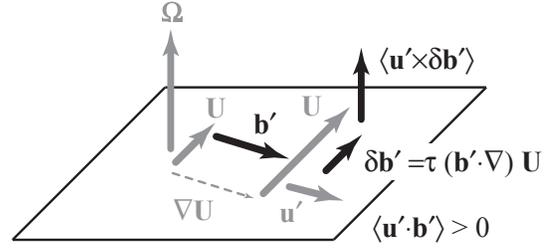}
\caption{\label{fig:emf_u_times_delb} Magnetic fluctuation in the mean-flow inhomogeneity. Contribution to the turbulent electromotive force (EMF) is parallel (antiparallel) to the mean vorticity if the cross helicity is positive (negative) in turbulence.}
\end{center}
\end{figure}

\paragraph{Magnetic fluctuation in mean-flow shear}
	Let us consider a magnetic fluctuation ${\bf{b}}'$ located in a large-scale vortical or rotational motion (Fig.~\ref{fig:emf_u_times_delb}). Induction of the magnetic fluctuation arising from the mean fluid motion is subject to
\begin{equation}
	\frac{\partial {\bf{b}}'}{\partial t} 
	= \nabla \times \left( {{\bf{U}} \times {\bf{b}}'} \right)
	+ \cdots
	\label{eq:fluct_b_eq_U1}
\end{equation}
or equivalently
\begin{equation}
	\frac{D{\bf{b}}'}{Dt}
	= - {\bf{b}}' \nabla \cdot {\bf{U}}
	+ \left( {{\bf{b}}' \cdot \nabla} \right) {\bf{U}}
	+ \cdots
	\label{eq:fluct_b_eq_U2}
\end{equation}
[see also Eq.~(\ref{eq:fluct_b_eq})]. The first term in Eq.~(\ref{eq:fluct_b_eq_U2}) is the mean-flow dilatation effect, which vanishes in the incompressible or solenoidal velocity case. The second term is associated with the rate of change that is observed in the mean velocity ${\bf{U}}$ as we move along ${\bf{b}}'$. The $\delta {\bf{b}}'$ induction due to this effect is written as
\begin{equation}
	\delta {\bf{b}}' = \tau \left( {{\bf{b}}' \cdot \nabla} \right) {\bf{U}}.
	\label{eq:del_fluct_b_U}
\end{equation}
The direction of $\delta {\bf{b}}'$ is in the direction of the mean flow ${\bf{U}}$ as in Fig.~\ref{fig:emf_u_times_delb}. This effect may be regarded as the turbulent counterpart of the differential-rotation effect [Eq.~(\ref{eq:mag_adv})]. If we assume that the turbulent cross helicity is positive ($\langle {{\bf{u}}' \cdot {\bf{b}}'} \rangle > 0$), the velocity fluctuation ${\bf{u}}'$ is statistically aligned with the magnetic fluctuation ${\bf{b}}'$, although the relationship between the turbulent velocity and magnetic-field vectors is much more random in each realization. It follows from Eq.~(\ref{eq:del_fluct_b_U}) that we have a contribution to the turbulent EMF as $\langle {{\bf{u}}' \times \delta{\bf{b}}'} \rangle$, which is parallel to the mean vorticity $\mbox{\boldmath$\Omega$}$ for a positive turbulent cross helicity (Fig.~\ref{fig:emf_u_times_delb}). For a negative turbulent cross helicity ($\langle {{\bf{u}}' \cdot {\bf{b}}'} \rangle < 0$), the contribution is antiparallel to $\mbox{\boldmath$\Omega$}$. In Fig.~\ref{fig:emf_u_times_delb}, only a case with increasing ${\bf{U}}$ along ${\bf{b}}'$ is depicted. In the opposite case with decreasing ${\bf{U}}$ along ${\bf{b}}'$, the direction of $\delta {\bf{b}}'$ is antiparallel to the mean velocity ${\bf{U}}$. Even in this case, the final contribution to the turbulent EMF, $\langle {{\bf{u}}' \times \delta {\bf{b}}'} \rangle$, is parallel to the mean vorticity $\mbox{\boldmath$\Omega$}$ as long as the turbulent cross helicity is positive.

	 The above argument shows that the combination of the modulation of the magnetic fluctuation through the inhomogeneous mean flow and the cross-correlation between the velocity and magnetic fluctuations leads to the cross-helicity effect in the turbulent EMF.

\begin{figure}[htb]
\begin{center}
\includegraphics[width=.40\textwidth]{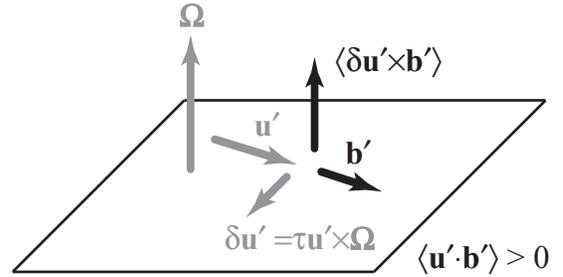}
\caption{\label{fig:emf_delu_times_b} Velocity fluctuation in the mean-flow inhomogeneity. Contribution to the turbulent electromotive force (EMF) is parallel to the mean vorticity if the cross helicity is positive in turbulence.}
\end{center}
\end{figure}

\paragraph{Velocity fluctuation in mean-flow shear}
We consider a fluid element fluctuating (${\bf{u}}'$) in the mean vortical or rotational motion $\mbox{\boldmath$\Omega$}$ shown in Fig.~\ref{fig:emf_delu_times_b}. The motion is subject to the mean vortical motion, which is locally equivalent to the rotation with an angular velocity of $\mbox{\boldmath$\omega$}_{\rm{F}} = \mbox{\boldmath$\Omega$}/2$. A modulation of the velocity fluctuation, $\delta{\bf{u}}'$, is induced by the Colioris-like force as
\begin{equation}
	\delta {\bf{u}}'
	= \tau {\bf{u}}' \times \mbox{\boldmath$\Omega$}.
	\label{eq:del_u_in_Omega}
\end{equation} 
As in the previous case, we assume a positive cross helicity in turbulence ($\langle {{\bf{u}}' \cdot {\bf{b}}'} \rangle > 0$). Due to this assumption, the magnetic fluctuation ${\bf{b}}'$ is statistically aligned with the velocity fluctuation ${\bf{u}}'$. It follows from Eq.~(\ref{eq:del_u_in_Omega}) that we have a contribution to the turbulent EMF as $\langle {\delta {\bf{u}}' \times {\bf{b}}'} \rangle$, which is parallel to the mean vorticity $\mbox{\boldmath$\Omega$}$ for a positive turbulent cross helicity (Fig.~\ref{fig:emf_delu_times_b}).

	The above argument shows that it is the combination of the local angular-momentum conservation and the cross helicity in turbulence that induces the cross-helicity effect in the turbulent EMF.

\subsubsection{Field configurations}
	In the usual dynamo model, the turbulent EMF is expressed by Eq.~(\ref{eq:emf_model_full}) with the cross-helicity-related term $\gamma \mbox{\boldmath$\Omega$}$ dropped. There the turbulent EMF is constituted of the turbulent magnetic diffusivity $\beta$ and the helicity or $\alpha$ effects. The turbulent magnetic diffusivity $\beta$ always arises as far as turbulence is present. In this sense, $\beta$ is the primary effect of turbulence in the mean magnetic-field evolution. If the mirrorsymmetry of turbulence is broken typically by a rotation, some other turbulence effects such as the $\alpha$ effect enter into the turbulent EMF. In case the $\beta$ effect is mainly balanced by the $\alpha$ effect, a mean-field configuration with the alignment between the mean electric-current density and magnetic field, ${\bf{J}}$ and ${\bf{B}}$, is expected. In an extreme situation, the so-called force-free field configuration (${\bf{B}} \parallel {\bf{J}}$, ${\bf{J}} \times {\bf{B}}= 0$) arises. The direction of ${\bf{B}}$ relative to ${\bf{J}}$ (parallel or antiparallel) depends on the sign of $\alpha$. In the other extreme case where the $\beta$ effect is mainly balanced by the cross-helicity or $\gamma$ effect, a mean-field configuration with the alignment between the mean electric-current density and vorticity, ${\bf{J}}$ and $\mbox{\boldmath$\Omega$}$, is expected. One possible configuration that leads to this alignment of ${\bf{J}}$ and $\mbox{\boldmath$\Omega$}$ is the alignment of ${\bf{B}}$ with the mean flow ${\bf{U}}$. This suggests that a typical mean-field configuration in the cross-helicity dynamo is the mean magnetic field aligned to the mean velocity.

	These situations may be schematically expressed as
\begin{equation}
	\overunderbraces{&\br{2}{\alpha\ \mbox{dynamo}}}%
	{{\bf{E}}_{\rm{M}} 
	 = &\alpha {\bf{B}} &- \beta {\bf{J}}&
	 +&\gamma \mbox{\boldmath$\Omega$}}%
	 {& &\br{3}{\mbox{cross-helicity dynamo}}}.
	 \label{eq:alpha_crhel_dynamo}
\end{equation}
Which of the $\alpha$ and $\gamma$ effects plays a more relevant role in turbulent dynamo depends on how much $\alpha$ and $\gamma$ we have in turbulence in addition to how much strong mean magnetic field and vorticity, ${\bf{B}}$ and $\mbox{\boldmath$\Omega$}$, we have. We already saw in Eq.~(\ref{eq:abc_model}) that the transport coefficients are directly related to the statistical property of turbulence. From the turbulent dynamo viewpoint, the mean-field configurations realized in astrophysical phenomena are determined by how much residual and cross helicities we have in turbulence and how they are spatiotemporally distributed in relation to the turbulent energy, which determines the dynamics of the turbulent magnetic diffusivity $\beta$.

	It is worthwhile to note that, the turbulent EMF expression [Eq.~(\ref{eq:emf_model_full})] was validated with the aid of a direct numerical simulation (DNS) of the Kolmogorov flow with an imposed magnetic field in the inhomogeneous direction \citep{yok2011a}. In the numerical experiment, it was shown that the cross-helicity effect coupled with the mean vorticity is the main balancer to the turbulent magnetic diffusivity, and the role of the helicity effect in this particular flow is almost negligible. For detailed explanations of the cross helicity and related dynamo, the reader is referred to \cite{yok2013a}.

\subsubsection{Cross-helicity evolution}
	Thus far, we have seen that the turbulent electromotive force (EMF) has a contribution from the mean-flow inhomogeneity in the presence of the turbulent cross helicity. How much cross helicity we have in turbulence is another problem. In this subsection, we address this problem by considering the transport equation of the turbulent cross helicity.

	From Eqs.~(\ref{eq:fluct_u_eq}) and (\ref{eq:fluct_b_eq}), the evolution equation of the turbulent cross helicity $W$ is written as
\begin{equation}
	\frac{DW}{Dt}
	= - {\cal{R}}^{ab} \frac{\partial B^b}{\partial x^a}
	- {\bf{E}}_{\rm{M}} \cdot \mbox{\bf{\boldmath$\Omega$}}
	- \varepsilon_W
	+ T_W,
	\label{eq:W_eq}
\end{equation}
where $\mbox{\boldmath${\cal{R}}$} = \{ {{\cal{R}}^{\alpha\beta}} \}$ is the Reynolds stress defined by
\begin{equation}
	{\cal{R}}^{\alpha\beta}
	= \langle {u'{}^\alpha u'{}^\beta - b'{}^\alpha b'{}^\beta} \rangle.
	\label{eq:Rey_strss_def}
\end{equation}
Strictly speaking, this should be called the Reynolds stress subtracted by the turbulent Maxwell stress, but we denote Eq.~(\ref{eq:Rey_strss_def}) as the Reynolds stress of MHD turbulence in this work.

	In Eq.~(\ref{eq:W_eq}), the first and the second terms are called the production terms since they express how much cross helicity is generated in turbulence through the cascade from the mean-field cross helicity ${\bf{U}} \cdot {\bf{B}}$. This point is clearer if we write the evolution equation of ${\bf{U}} \cdot {\bf{B}}$ as
\begin{equation}
	\frac{D}{Dt} {\bf{U}} \cdot {\bf{B}}
	= + {\cal{R}}^{ab} \frac{\partial B^b}{\partial x^a}
	+ {\bf{E}}_{\rm{M}} \cdot \mbox{\bf{\boldmath$\Omega$}}
	+ \cdots,
	\label{eq:mean_cr-hel_eq}
\end{equation}
which contains exactly the same terms but with the opposite signs as Eq.~(\ref{eq:W_eq}).

	The third and fourth terms in Eq.~(\ref{eq:W_eq}), $\varepsilon_W$ and $T_W$, are the dissipation and transport terms, respectively, whose definitions and detailed expressions are suppressed here. In turbulence modeling, the transport terms are expressed as
\begin{equation}
	T_W
	= {\bf{B}} \cdot \nabla K 
	+ \nabla \cdot \left( {\frac{\nu_{\rm{K}}}{\sigma_W} \nabla W} \right),
	\label{eq:T_W_model}
\end{equation}
where $\nu_{\rm{K}}$ is the turbulent viscosity and $\sigma_W$ is the Prandtl number for the cross-helicity diffusion.

	For detailed arguments on the turbulent cross-helicity evolution including the modeling of the turbulent cross-helicity dissipation rate $\varepsilon_W$, the reader is referred to \cite{yok2011b}.

\section{Dynamo equations with the cross-helicity effect\label{sec:3}}
\subsection{Basic dynamo equations with the cross-helicity effect}
	If we adopt Eq.~(\ref{eq:emf_model_full}) for ${\bf{E}}_{\rm{M}}$, the mean magnetic induction equation is written as
\begin{equation}
	\frac{\partial{\bf{B}}}{\partial t} 
	= \nabla \times ({\bf{U}} \times {\bf{B}})
	+ \nabla \times \left( {
	- \beta \nabla \times {\bf{B}}
	+ \alpha {\bf{B}}
	+ \gamma \mbox{\boldmath$\Omega$}
	} \right).
	\label{eq:dynamo_B_eq}
\end{equation}

	In usual simple dynamo models such as the Parker model, only the transport equations for the magnetic field are taken into account with the transport coefficients being treated as parameters. In contrast, in the present work, in addition to the ${\bf{B}}$ equation, we consider the spatiotemporal evolution of the transport coefficient $\gamma$, which is related to the cross-helicity effect. On the basis of Eq.~(\ref{eq:W_eq}), the $\gamma$ equation is modeled as
\begin{equation}
	\frac{\partial \gamma}{\partial t}
	= \beta \nabla^2 \gamma
	- \alpha \tau {\bf{B}} \cdot \mbox{\boldmath$\Omega$}
	+ \beta \tau (\nabla \times {\bf{B}}) \cdot \mbox{\boldmath$\Omega$}
	- \gamma \tau \mbox{\boldmath$\Omega$}^2.
	\label{eq:dynamo_gamma_eq}
\end{equation}
Here, the first term arises from the second term in Eq.~(\ref{eq:T_W_model}), and the second to fourth terms come from the second term in Eq.~(\ref{eq:W_eq}) with ${\bf{E}}_{\rm{M}}$ [Eq.~(\ref{eq:emf_model_full})].

	Equations~(\ref{eq:dynamo_B_eq}) and (\ref{eq:dynamo_gamma_eq}) constitute the basic dynamo equations with the cross-helicity effect.

\subsection{Assumptions on the rotation and magnetic field}
	Let us consider the local Cartesian coordinate $(x,y,z)$, where $x$, $y$, and $z$ axes are directed in the colatitudinal ($\theta$), azimuthal ($\phi$), and radial ($r$) directions, respectively:
\begin{equation}
	(x,y,z) = (\theta,\phi,r).
	\label{eq:local_cartesian}
\end{equation}

	We assume the azimuthal symmetry:
\begin{equation}
	{\partial}/{\partial \phi} = {\partial}/{\partial y} = 0.
	\label{eq:azimuthal_sym}
\end{equation}

	As for the solar rotation in the convection zone, we assume that the mean velocity is only in the azimuthal direction and omit the meridional circulations as
\begin{equation}
	{\bf{U}} = (U^x, U^y, U^z) = (0,U^y,0),
	\label{eq:mean_vel_form}
\end{equation}
with the mean-velocity shear in the latitudinal and radial direction, $\partial U^y / \partial x [\sim (1/r) \partial U^\phi / \partial \theta]$ and $\partial U^y / \partial z (\sim \partial U^\phi / \partial r)$.

	As usual, the mean magnetic field is decomposed into the toroidal and poloidal components, ${\bf{B}}_{\rm{tor}}$ and ${\bf{B}}_{\rm{pol}}$, respectively as
\begin{eqnarray}
	{\bf{B}} 
	&=& (B^x, B^y, B^z) = {\bf{B}}_{\rm{tor}} + {\bf{B}}_{\rm{pol}}
	\nonumber\\
	&=& (0, B^y, 0) + \nabla \times (0, A^y, 0)
	\nonumber\\
	&=& (0, B^y, 0) 
	+ \left( {
		- \frac{\partial A^y}{\partial z}, 0, \frac{\partial A^y}{\partial x}
	} \right).
	\label{eq:mean_mag_form}
\end{eqnarray}
Hereafter we drop index $y$ on $B^y$, $A^y$, and $U^y$ for the brevity of notation.

\subsection{Basic dynamo equations}
	We further omit the $\alpha$-related term in the equation for $B$. This is because we assume the toroidal magnetic field is dominantly generated by the mean-velocity inhomogeneities. Under these assumptions, Eqs.~(\ref{eq:dynamo_B_eq}) and (\ref{eq:dynamo_gamma_eq}) are reduced to the following equations.
	
\noindent{\it{Vector potential for poloidal field}}:
\begin{equation}
	\frac{\partial A}{\partial t}
	= \beta \left( {
		\frac{\partial^2 A}{\partial x^2}
		+ \frac{\partial^2A}{\partial z^2}
	} \right)
	+ \alpha B,
	\label{eq:basic_pol_eq}
\end{equation}
\noindent{\it{Toloidal field}}:
\begin{eqnarray}
	\frac{\partial B}{\partial t}
	&=& \beta \left( {
		\frac{\partial^2 B}{\partial x^2}
		+ \frac{\partial^2 B}{\partial z^2}
	} \right)
	- \frac{\partial^2 U}{\partial z^2} \gamma
	- \frac{\partial^2 U}{\partial x^2} \gamma
	\nonumber\\
	& &\hspace{-10pt} - \frac{\partial U}{\partial z}\frac{\partial \gamma}{\partial z}
	- \frac{\partial U}{\partial x}\frac{\partial \gamma}{\partial x}
	- \frac{\partial U}{\partial z}\frac{\partial A}{\partial x}
	- \frac{\partial U}{\partial x}\frac{\partial A}{\partial z},
	\label{eq:basic_tor_eq}
\end{eqnarray}
\noindent{\it{Cross helicity}}:
\begin{eqnarray}
	\frac{\partial \gamma}{\partial t}
	&=& \beta \left( {
		\frac{\partial^2 \gamma}{\partial x^2}
		+ \frac{\partial^2 \gamma}{\partial z^2}
	} \right)
	- \alpha \tau \left( {
		\frac{\partial U}{\partial x} \frac{\partial A}{\partial x}
		+ \frac{\partial U}{\partial z} \frac{\partial A}{\partial z}
	} \right)
	\nonumber\\
	& & + \beta \tau \left( {
		\frac{\partial U}{\partial x} \frac{\partial B}{\partial x}
		- \frac{\partial U}{\partial z} \frac{\partial B}{\partial z}
	} \right)
	\nonumber\\
	& & - \gamma \tau \left[ {
		\left( {\frac{\partial U}{\partial x}} \right)^2
		+ \left( {\frac{\partial U}{\partial z}} \right)^2
	} \right].
	\label{eq:basic_gamma_eq}
\end{eqnarray}

\subsection{Non-dimensionalization}
	We introduce non-dimensional variables as
\begin{eqnarray}
	&&B=B_0 \tilde{B},\; 
	A=B_0 L \tilde{A},\; 
	\gamma = \gamma_0 \tilde{\gamma} = B_0 L \tilde{\gamma},\; 
	x = L \tilde{x},
	\nonumber\\	
	&&t = ({L^2}/{\beta_0}) \tilde{t},\;
	U = U_0 \tilde {U},\; 
	\alpha = \alpha_0 \tilde{\alpha},\;
	\beta = \beta_0 \tilde{\beta}
	\label{eq:non-dim_vars}
\end{eqnarray}
with characteristic magnetic-field strength $B_0$, length scale $L$, cross-helicity effect $\gamma_0$, turbulent magnetic diffusivity $\beta_0$, mean-flow speed $U_0$, helicity effect $\alpha_0$.

	We assume some symmetries with respect to the equator ($x=\pi/2$) for the mean-flow speed $U$, helicity effect $\alpha$, turbulent magnetic diffusivity $\beta$, and time scale $t$ as
\begin{equation}
	\tilde{U} = \sin x,\; \tilde{\alpha} = \cos x,\; \tilde{\beta} = 1,\; \tilde{t} =1.
	\label{eq:sym_equator}
\end{equation}
Namely, at the equator, the mean flow is symmetric and maximum while the helicity is antisymmetric and vanishes. The turbulent magnetic diffusivity and characteristic time scale are uniform.

	We omit all the tilde $\tilde{}$ of the non-dimensional variables in the following. As for the radial or $z$ dependence of $A$, $B$, and $\gamma$, we assume $\exp(ikz)$ dependence, and assume $\partial U / \partial z = k_u U$. Then Eqs.~(\ref{eq:basic_pol_eq})-(\ref{eq:basic_gamma_eq}) are rewritten as
\begin{equation}
	\frac{\partial A}{\partial t} 
	= \frac{\partial^2 A}{\partial x^2}
	- k^2 A
	+ R_\alpha \cos x B,
	\label{eq:non-d_pol_eq}
\end{equation}
\begin{eqnarray}
	\frac{\partial B}{\partial t} 
	&=& \frac{\partial^2 B}{\partial x^2}
	- k^2 B
	\nonumber\\
	&-& R_u \left( {
		k_u^2 \sin x \gamma + i k k_u \sin x \gamma
		- \sin x \gamma
		+ \cos x \frac{\partial \gamma}{\partial x}
	} \right)
	\nonumber\\
	&+& R_u \left( {
		k_u \sin x \frac{\partial A}{\partial x} - i k \cos x A
	} \right),
	\label{eq:non-d_tor_eq}
\end{eqnarray}
\begin{eqnarray}
	\frac{\partial \gamma}{\partial t}
	&=& \frac{\partial^2 \gamma}{\partial x^2}
	- k^2 \gamma
	\nonumber\\
	&-& R_\alpha R_u \left( {
		\cos^2 x \frac{\partial A}{\partial x} - i k k_u \sin x \cos x A
	} \right)
	\nonumber\\ 
	&+& R_u \left( {
		\cos x \frac{\partial B}{\partial x} - i k k_u \sin x B
	} \right)
	\nonumber\\
	&-& R_u^2 \left( {
		\cos^2 x + k_u^2 \sin^2 x
	} \right) \gamma,
	\label{eq:non-d_gamma_eq}
\end{eqnarray}
where
\begin{equation}
	R_u = {U_0 L}/{\beta_0},
	\label{eq:turb_Re_def}
\end{equation}
\begin{equation}
	R_\alpha = {\alpha_0 L}/{\beta_0}.
	\label{eq:alpha_Re_def}
\end{equation}
The first one $R_u$ is the turbulent Reynolds number based on the turbulent magnetic diffusivity $\beta$. The second one $R_\alpha$ represents the relative amplitude of the helicity or $\alpha$ effect to the turbulent magnetic diffusivity $\beta$. These two non-dimensional numbers determine the basic behavior of the magnetic field and the cross helicity in this system of equations.

	For the sake of simplicity, we assume $k=0$, i.e., neglect the radial or $z$ derivatives of $A$, $B$, and $\gamma$. It implies from Eq.~(\ref{eq:mean_mag_form}) that the colatitudinal component of the magnetic field vanishes ($B^x \simeq B^\theta = 0$).
	
	Here, we introduce $f_i = \{ {0,\; 1} \}$ to switch individual terms on and off. With $f_i$ Eqs.~(\ref{eq:non-d_pol_eq})-(\ref{eq:non-d_gamma_eq}) are written as
\begin{equation}
	\frac{\partial A}{\partial t} 
	= \frac{\partial^2 A}{\partial x^2}
	+ R_\alpha \cos x B,
	\label{eq:on-off_pol_eq}
\end{equation}
\begin{eqnarray}
	\frac{\partial B}{\partial t} 
	&=& \frac{\partial^2 B}{\partial x^2}
	+ R_u \left( {
		- f_1 k_u^2 \sin x \gamma
		+ f_2 \sin x \gamma
	\rule{0.ex}{3.ex}} \right.
	\nonumber\\
	& & \left. {
		- f_3 \cos x \frac{\partial \gamma}{\partial x}
	} \right)
	+ R_u \left( {
		f_4 k_u \sin x \frac{\partial A}{\partial x}
	} \right),
	\label{eq:on-off_tor_eq}
\end{eqnarray}
\begin{eqnarray}
	\frac{\partial \gamma}{\partial t}
	&=& \frac{\partial^2 \gamma}{\partial x^2}
	- R_\alpha R_u f_5 \left( {
		\cos^2 x \frac{\partial A}{\partial x}
	} \right)
	+ R_u f_6 \cos x \frac{\partial B}{\partial x}
	\nonumber\\
	& & - R_u^2 f_7 \left( {
		\cos^2 x + k_u^2 \sin^2 x
	} \right) \gamma.
	\label{eq:on-off_gamma_eq}
\end{eqnarray}

	If we put $f_1 = f_2 = f_3 = f_5 = f_6 = f_7 = 0, f_4 = 1$ in Eqs.~(\ref{eq:on-off_pol_eq})-(\ref{eq:on-off_gamma_eq}) and omit the equation of $\gamma$ itself, we get the usual $\alpha$--$\Omega$ dynamo model:
\begin{equation}
	\frac{\partial A}{\partial t} 
	= \frac{\partial^2 A}{\partial x^2}
	+ R_\alpha \cos x B,
	\label{eq:alpha-Omega_pol_eq}
\end{equation}
\begin{eqnarray}
	\frac{\partial B}{\partial t} 
	&=& \frac{\partial^2 B}{\partial x^2}
	+ R_u \left( {
		k_u \sin x \frac{\partial A}{\partial x}
	} \right).
	\label{eq:alpha-Omega_tor_eq}
\end{eqnarray}

	If we put $f_1 = f_3 = f_4 = f_6 = f_7 = 0, f_2 = f_5 = 1$, we reproduce the original cross-helicity dynamo model:
\begin{equation}
	\frac{\partial A}{\partial t} 
	= \frac{\partial^2 A}{\partial x^2}
	+ R_\alpha \cos x B,
	\label{eq:original_pol_eq}
\end{equation}
\begin{eqnarray}
	\frac{\partial B}{\partial t} 
	&=& \frac{\partial^2 B}{\partial x^2}
	+ R_u \sin x \gamma,
	\label{eq:original_tor_eq}
\end{eqnarray}
\begin{eqnarray}
	\frac{\partial \gamma}{\partial t}
	&=& \frac{\partial^2 \gamma}{\partial x^2}
	- R_\alpha R_u \left( {
		\cos^2 x \frac{\partial A}{\partial x}
	} \right).
	\label{eq:original_gamma_eq}
\end{eqnarray}

	We further introduce new variables
\begin{equation}
	\tilde{A} = R_\alpha R_u A,\;
	\tilde{B} = R_\alpha^2 R_u B,
\end{equation}
and omit the tilde $\tilde{}$ again, then we have
\begin{equation}
	\frac{\partial A}{\partial t} 
	= \frac{\partial^2 A}{\partial x^2}
	+ \cos x B,
	\label{eq:final_pol_eq}
\end{equation}
\begin{equation}
	\frac{\partial B}{\partial t} 
	= \frac{\partial^2 B}{\partial x^2}
	+ P^2 \sin x \gamma,
	\label{eq:final_tor_eq}
\end{equation}
\begin{equation}
	\frac{\partial \gamma}{\partial t}
	= \frac{\partial^2 \gamma}{\partial x^2}
	- \cos^2 x \frac{\partial A}{\partial x},
	\label{eq:final_gamma_eq}
\end{equation}
where
\begin{equation}
	P = R_\alpha R_u.
	\label{eq:dynamo_number_P_def}
\end{equation}
From Eqs.~(\ref{eq:final_pol_eq})-(\ref{eq:final_gamma_eq}) show that the solution depends only on the square of dynamo number $P = R_\alpha R_u$.

\subsection{Boundary conditions and free-decay modes}
	We put the characteristic length scale $L = \pi/2$. For the $x$ or colatitude coordinate, $x=0$, $x=\pi/2$, and $x=\pi$ correspond to the ``Northpole'', ``Equator'', and ``Southpole'', respectively. 

\begin{figure}[htb]
\begin{center}
\includegraphics[width=.40\textwidth]{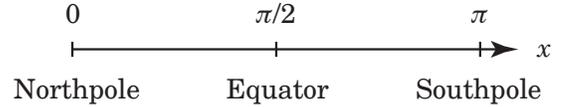}
\caption{\label{fig:colatitude_coord} Colatitudinal coordinate or $x$. $x=0$ corresponds to the northpole, $x=\pi/2$ the equator, $x=\pi$ the southpole.}
\end{center}
\end{figure}

	The boundary conditions for the solutions of $A$, $B$, and $\gamma$ for the poles should be
\begin{subequations}
\begin{equation}
	x = 0,\pi:\;\;\; 
	A = B =\frac{\partial \gamma}{\partial x} = 0.
	\label{eq:bc_poles}
\end{equation}
As for the equator, the boundary conditions for the antisymmetric solution (``dipole'') are
\begin{equation}
	x = \pi/2:\;\;\; 
	\frac{\partial A}{\partial x} = B = \gamma = 0,
	\label{eq:bc_eqtr_anti}
\end{equation}
and the ones for the symmetric solution (``quadruple'') are
\begin{equation}
	x = {\pi}/{2}:\;\;\; 
	A = \frac{\partial B}{\partial x} = \frac{\partial \gamma}{\partial x} = 0.
	\label{eq:bc_eqtr_sym}
\end{equation}
\end{subequations}
In what follows, we adopt only the boundary conditions for antisymmetric (``dipole'') solution.

	The boundary conditions for $\gamma$ or equivalently the turbulent cross helicity are not so clear as compared with those for $A$ and $B$ or the mean magnetic fields. In the dipole field configuration, from the antisymmetric property of the turbulent cross helicity, $\gamma$ should vanish at the equator as Eq.~(\ref{eq:bc_eqtr_anti}). However, the spatiotemporal distributions of the turbulent cross helicity near the poles are still open. Since we adopted a slab geometry with the local Cartesian coordinate Eq.~(\ref{eq:local_cartesian}), there are several differences between the boundary regions ($x=0, \pi$) and the real pole regions. With this point in mind, we adopt Eq.~(\ref{eq:bc_poles}) as the simplest boundary conditions for $\gamma$ or the turbulent cross helicity.

	We consider the case of ``free-decay'' of $A$, $B$, and $\gamma$, which obey 
\begin{equation}
	\frac{\partial A}{\partial t}
	= \frac{\partial^2 A}{\partial x^2},\;\;\;
	\frac{\partial B}{\partial t}
	= \frac{\partial^2 B}{\partial x^2},\;\;\;
	\frac{\partial \gamma}{\partial t}
	= \frac{\partial^2 \gamma}{\partial x^2}.
	\label{eq:free_decay_modes}
\end{equation}
In this situation, all of $A$, $B$, and $\gamma$ just decay due to the turbulent magnetic diffusion represented by the first terms of Eqs.~(\ref{eq:basic_pol_eq})-(\ref{eq:basic_gamma_eq}). The solutions of these equations, satisfying the boundary conditions Eqs.~(\ref{eq:bc_poles}) and (\ref{eq:bc_eqtr_anti}), can be written as
\begin{subequations}\label{eq:decay_modes}
\begin{equation}
	A_n = e^{\omega_n t} \sin nx\;\;\;
	\mbox{with}\;\;\;
	\omega_n = -n^2,\;\;\;
	n=1, 3, 5, \cdots,
	\label{eq:A_decay_modes}
\end{equation}
\begin{equation}
	B_n = e^{\omega_n t} \sin nx\;\;\;
	\mbox{with}\;\;\;
	\omega_n = -n^2,\;\;\;
	n=2, 4, 6, \cdots,
	\label{eq:B_decay_modes}
\end{equation}
\begin{equation}
	\gamma_n = e^{\omega_n t} \cos nx\;\;\;
	\mbox{with}\;\;\;
	\omega_n = -n^2,\;\;\;
	n=1, 3, 5, \cdots.
	\label{eq:gamma_decay_modes}
\end{equation}
\end{subequations}
Note that these solutions constitute a complete and orthogonal system of functions which already satisfies the boundary conditions [Eqs.~(\ref{eq:bc_poles}) and (\ref{eq:bc_eqtr_anti})].

\section{Eigenvalue analysis\label{sec:4}}
	Now we consider the solution of the present dynamo equations [Eqs.~(\ref{eq:final_pol_eq})-(\ref{eq:final_gamma_eq})] as eigenvalue problem. We may expand the dynamo solution in any complete orthogonal system of functions which satisfies the boundary conditions. For the sake of simplicity, here we expand the dynamo solution in the free-decay modes [Eq.~(\ref{eq:decay_modes})].
\begin{subequations}\label{eq:sol_decay_exp}
\begin{equation}
	A(x,t)
	= e^{\omega t} \sum_{n=1,3,5,\cdots}^{N-1} a_n \sin nx,
	\label{eq:A_decay_exp}
\end{equation}
\begin{equation}
	B(x,t)
	= e^{\omega t} \sum_{n=2,4,6,\cdots}^{N} b_n \sin nx,
	\label{eq:B_decay_exp}
\end{equation}
\begin{equation}
	\gamma(x,t)
	= e^{\omega t} \sum_{n=1,3,5,\cdots}^{N-1} c_n \cos nx.
	\label{eq:gamma_decay_exp}
\end{equation}
\end{subequations}

	Substituting Eq.~(\ref{eq:sol_decay_exp}) into Eqs.~(\ref{eq:final_pol_eq})-(\ref{eq:final_gamma_eq}), we obtain
\begin{eqnarray}
	\omega \sum_{n=1,3,\cdots} a_n \sin nx
	&=& - \sum_{n=1,3,\cdots} a_n n^2 \sin nx
	\nonumber\\
	&+& \cos x \sum_{n=2,4,\cdot} b_n \sin nx,
	\label{eq:a_n_eq}
\end{eqnarray}
\begin{eqnarray}
	\omega \sum_{n=2,4,\cdots} b_n \sin nx
	&=& - \sum_{n=2,4,\cdots} b_n n^2 \sin nx
	\nonumber\\
	&+& P^2 
		\sin x \sum_{n=1,3,\cdot} c_n \cos nx,
	\label{eq:b_n_eq}
\end{eqnarray}
\begin{eqnarray}
	\omega \sum_{n=1,3,\cdots} c_n \cos nx
	&=& - \sum_{n=1,3,\cdots} c_n n^2 \cos nx
	\nonumber\\
	&-& \cos^2 x \sum_{n=1,3,\cdot} a_n n \cos nx.
	\label{eq:c_n_eq}
\end{eqnarray}

	After simplifying Eqs.~(\ref{eq:a_n_eq})-(\ref{eq:c_n_eq}) with trigonometric relations, we multiply equations for $a_n$ and $b_n$ by $\sin mx$ and equation for $c_n$ by $\cos mx$, and integrate $(\pi/4) \int_0^{\pi/2} dx$. Using orthogonality relations such as
\begin{subequations}
\begin{equation}
	\int_0^{\pi/2} \sin nx \sin mx dx
	= \int_0^{\pi/2} \cos nx \cos mx
	= \frac{\pi}{4} \delta_{nm},
\end{equation}
\begin{equation}
	\int_0^{\pi/2} \sin nx \cos mx dx = 0,
\end{equation}
\end{subequations}
we obtain 
\begin{eqnarray}
	\omega a_m
	&=& - m^2 a_m
	+ \frac{1}{2} \left( {b_{m-1} + b_{m+1}} \right)
	\nonumber\\
	& & \hspace{50pt} (m = 1,3,5, \cdots, N),
\end{eqnarray}
\begin{eqnarray}
	\omega b_m
	&=& - m^2 b_m
	+ \frac{P^2}{2} \left( {c_{m-1} - c_{m+1}} \right)
	\nonumber\\
	&& \hspace{65pt} (m = 2,4,6, \cdots, N+1),
\end{eqnarray}
\begin{eqnarray}
	\omega c_m
	&=& -m^2 c_m
	\nonumber\\
	&& \hspace{-20pt} - \frac{1}{4} \left[ {
		(m-2) a_{m-2} + 2 m a_m + (m+2) a_{m+2}
	} \right]
	\nonumber\\
	&& \hspace{62pt} (m = 1,3,5, \cdots, N).
\end{eqnarray}

	We then have a matrix eigenvalue problem
\begin{equation}
	\omega {\bf{u}} = \mbox{\boldmath${\cal{M}}$} {\bf{u}},
	\label{eq:eigenvalue_problem}
\end{equation}
with the vector constituted by the expansion coefficients
\begin{equation}
	{\bf{u}} = \left( {
	a_1, b_2, c_1, a_3, b_4, c_3, a_5, b_6, c_5, \cdots, a_N, b_{N+1}, c_N
	} \right)^{\rm{T}},
	\label{eq:u_vec_def}
\end{equation}
and the matrix
\begin{widetext}
\begin{equation}
{\mbox{\boldmath${\cal{M}}$}} = \left({ \begin{array}{*{20}c}
   { - 1} & {{1 \mathord{\left/
 {\vphantom {1 2}} \right.
 \kern-\nulldelimiterspace} 2}} & {} & {} & {} & {} & {} & {} & {} & {} & {} & {}  \\
   {} & { - 4} & {{{P^2 } \mathord{\left/
 {\vphantom {{P^2 } 2}} \right.
 \kern-\nulldelimiterspace} 2}} & {} & {} & {{{-P^2 } \mathord{\left/
 {\vphantom {{-P^2 } 2}} \right.
 \kern-\nulldelimiterspace} 2}} & {} & {} & {} & {} & {} & {}  \\
   {{{ - 2} \mathord{\left/
 {\vphantom {{ - 2} 4}} \right.
 \kern-\nulldelimiterspace} 4}} & {} & { - 1} & {{{ - 3} \mathord{\left/
 {\vphantom {{ - 3} 4}} \right.
 \kern-\nulldelimiterspace} 4}} & {} & {} & {} & {} & {} & {} & {} & {}  \\
   {} & {{1 \mathord{\left/
 {\vphantom {1 2}} \right.
 \kern-\nulldelimiterspace} 2}} & {} & { - 9} & {{1 \mathord{\left/
 {\vphantom {1 2}} \right.
 \kern-\nulldelimiterspace} 2}} & {} & {} & {} & {} & {} & {} & {}  \\
   {} & {} & {} & {} & { - 16} & {{{P^2 } \mathord{\left/
 {\vphantom {{P^2 } 2}} \right.
 \kern-\nulldelimiterspace} 2}} & {} & {} & {{{-P^2 } \mathord{\left/
 {\vphantom {{-P^2 } 2}} \right.
 \kern-\nulldelimiterspace} 2}} & {} & {} & {}  \\
   {{{ - 1} \mathord{\left/
 {\vphantom {{ - 1} 4}} \right.
 \kern-\nulldelimiterspace} 4}} & {} & {} & {{{ - 6} \mathord{\left/
 {\vphantom {{ - 6} 4}} \right.
 \kern-\nulldelimiterspace} 4}} & {} & { - 9} & {{{ - 5} \mathord{\left/
 {\vphantom {{ - 5} 4}} \right.
 \kern-\nulldelimiterspace} 4}} & {} & {} & {} & {} & {}  \\
   {} & {} & {} & {} & {} & {} &  \ddots  & {} & {} & {} & {} & {}  \\
   {} & {} & {} & {} & {} & {} & {} &  \ddots  & {} & {} & {} & {}  \\
   {} & {} & {} & {} & {} & {} & {} & {} &  \ddots  & {} & {} & {}  \\
   {} & {} & {} & {} & {} & {} & {} & {{1 \mathord{\left/
 {\vphantom {1 2}} \right.
 \kern-\nulldelimiterspace} 2}} & {} & { - N^2 } & {{1 \mathord{\left/
 {\vphantom {1 2}} \right.
 \kern-\nulldelimiterspace} 2}} & {}  \\
   {} & {} & {} & {} & {} & {} & {} & {} & {} & {} & { - \left( {N + 1} \right)^2 } & {{{P^{2} } \mathord{\left/
 {\vphantom {{P^{2 \to } } 2}} \right.
 \kern-\nulldelimiterspace} 2}}  \\
   {} & {} & {} & {} & {} & {} & {{{ - \left( {N - 2} \right)} \mathord{\left/
 {\vphantom {{ - \left( {N - 2} \right)} 4}} \right.
 \kern-\nulldelimiterspace} 4}} & {} & {} & {{{ - 2N} \mathord{\left/
 {\vphantom {{ - 2N} 4}} \right.
 \kern-\nulldelimiterspace} 4}} & {} & { - N^2 }  \\
\end{array}} \right).
	\label{eq:matrix}
\end{equation}
\end{widetext}

\begin{figure}[htb]
\begin{center}
\includegraphics[width=.40\textwidth]{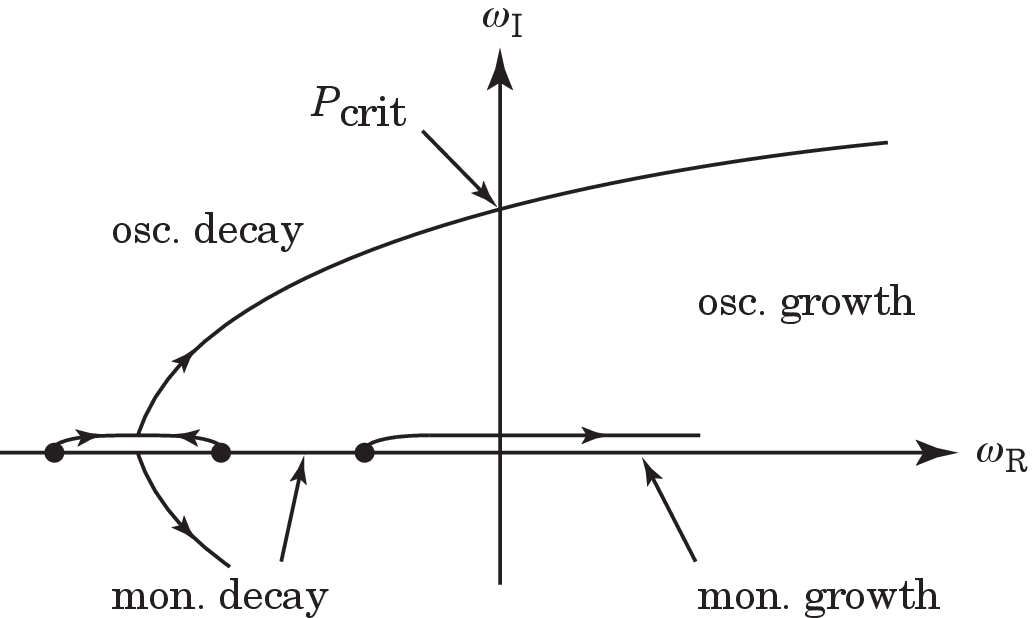}
\caption{\label{fig:omega_diagram} Schematic diagram of $\omega(P)$. }
\end{center}
\end{figure}

	The eigenvalue $\omega$ is expressed in terms of $P$ [Eq.~(\ref{eq:dynamo_number_P_def})]. We display $\omega(P)$ diagrams for $P>0$ and $P<0$. The diagram is schematically depicted in Fig.~{\ref{fig:omega_diagram}}. 

	The vanishing of $\omega_{\rm{I}}$ ($\omega_{\rm{I}} = 0$) gives the monotonic growth ($\omega_{\rm{R}} > 0$) or decay ($\omega_{\rm{R}} < 0$). The cases with a finite $\omega_{\rm{I}}$ correspond to the oscillational growth ($\omega_{\rm{R}} > 0$) or decay ($\omega_{\rm{R}} < 0$). The critical dynamo number is determined where $\omega_{\rm{R}} (P_{\rm{crit}}) = 0$ in Fig.~\ref{fig:omega_diagram}. This critical value $P_{\rm{crit}}$ gives the pure (neither growth nor decay) oscillation of the magnetic fields and the turbulent cross helicity.

\section{Butterfly diagrams\label{sec:5}}
\begin{figure*}[htb]
\begin{center}
\includegraphics[width=0.7\textwidth]{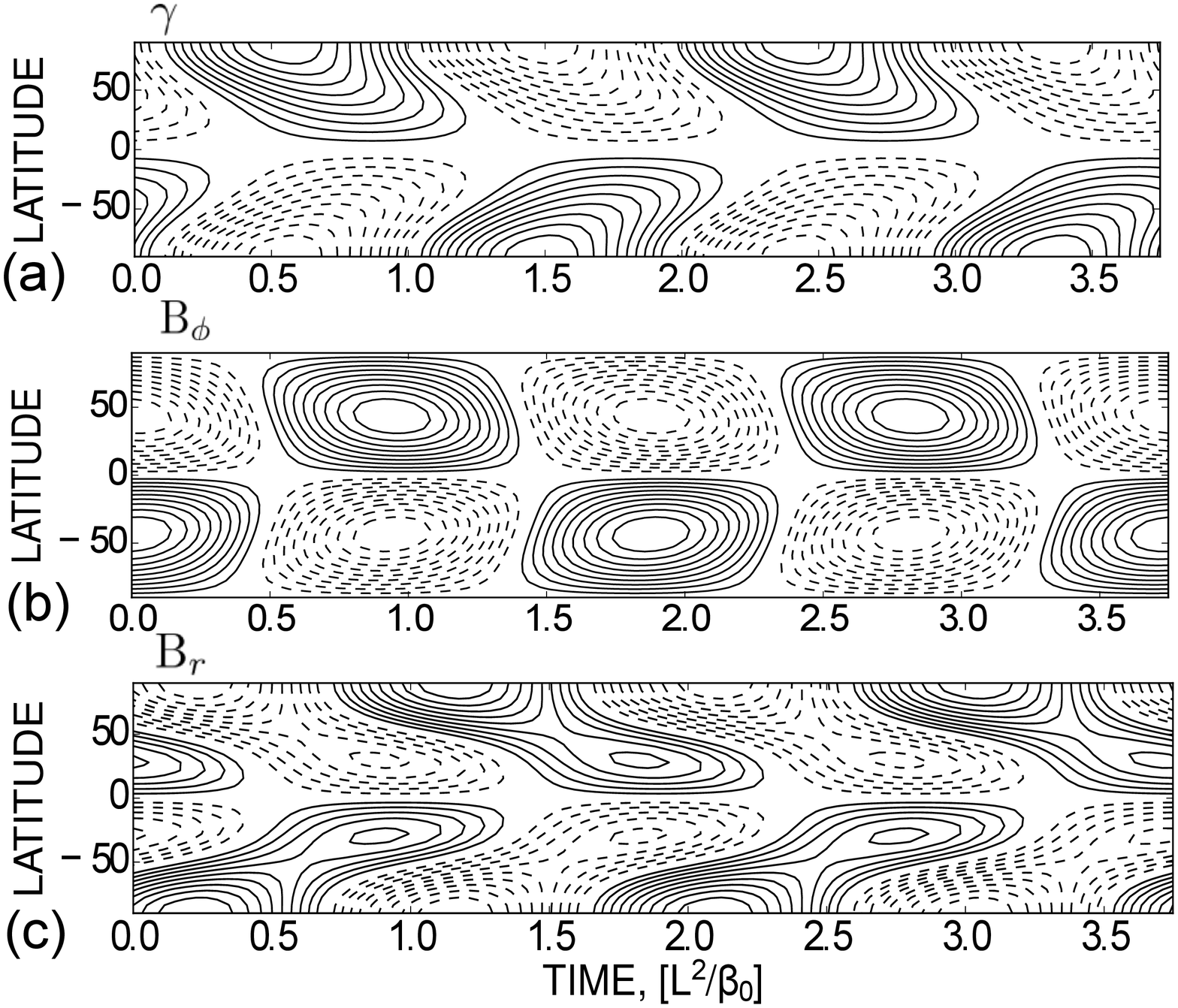}
\caption{\label{fig:butterfly_diagram} Butterfly diagram. (a) The cross-helicity related coefficient $\gamma\ [-0.5, 0.5]$. (b) The toroidal field $B^\phi = B^y\ [-0.8, 0.8]$. (c) The poloidal magnetic field $B^r = B^z (= \partial A / \partial x)\ [-0.2,0.2]$. The solid and dashed contours denote the positive and negative values, respectively. The equator (latitude $0^\circ$) corresponds to colatitude of $\theta = x = \pi / 2$.}
\end{center}
\end{figure*}

	An oscillational solution is marginally excited at
\begin{equation}
	P_{\rm{crit}} = 18.1\;\;\;
	\mbox{with}\;\;\;
	\omega_{\rm{I}} = 3.4.
	\label{eq:P_crit}
\end{equation}
	The butterfly diagrams of the magnetic fields and the turbulent cross helicity are shown in Fig.~\ref{fig:butterfly_diagram}, where the colatitudinal distribution of the poloidal magnetic field $B^r(\theta,t) = B^z(x,t) = \partial A / \partial x$ is plotted in color and the toroidal magnetic field $B^\phi(\theta,t) = B^y(x,t)$ in contour (top), and the turbulent cross-helicity-related coefficient $\gamma(x,t)$ (bottom) are displayed against time. 

	Here, the non-dimensional time is defined by Eq.~(\ref{eq:non-dim_vars}). The non-dimensional time interval $\hat{t} = 1$ corresponds to $t=45\ {\rm{yr}}$ for $\beta_0 = 10^{12}\ {\rm{cm}}^2\ {\rm{s}}^{-1}$ and $t=4.5\ {\rm{yr}}$ for $\beta_0 = 10^{13}\ {\rm{cm}}^2\ {\rm{s}}^{-1}$.
	
	We see from Fig.~\ref{fig:butterfly_diagram} (a) that, approximately from time $0$ to $1$, the positive and negative cross helicities are generated in the northern and southern hemispheres, respectively. The generation starts in the higher latitude regions and extends towards the lower latitude regions as the time proceeds.
	
	Figure~\ref{fig:butterfly_diagram} (b) shows that the induction of the mean toroidal magnetic field $B^y$ [positive (negative) in the northern (southern) hemisphere] follows the generation of the turbulent cross helicity $\gamma$. The induction is first prominent in the higher latitude regions then gradually moves to the lower latitude regions. 
	
	Figure~\ref{fig:butterfly_diagram} (c) shows that the generation of the poloidal magnetic field $B^z$ [positive (negative) in the northern (southern) hemisphere] follows that of the toroidal magnetic field $B^y$. At the initial stage of generation, the poloidal field is prominent in the higher latitude region then its dominant part moves to the lower latitude region.

	Once the poloidal magnetic field has been induced, the turbulent cross helicity with the opposite sign [negative (positive) in the northern (southern) hemisphere] starts being generated. This process starts in the higher latitude regions and moves to the lower latitude regions.

\section{Concluding remarks\label{sec:6}}
	We proposed a new simple dynamo model for the stellar activity cycle, in which the effect of the mean-flow inhomogeneity on turbulence is incorporated through the turbulent cross helicity (velocity--magnetic-field correlation). In addition to the mean magnetic-field equations, the transport equation of the turbulent cross helicity is simultaneously solved in the model. The basic scenario of this dynamo model is as follows.

\begin{itemize}

\item
The positive or negative turbulent cross helicity, whose total amount through the full volume or sphere is zero, is locally generated by the inhomogeneous configurations of the mean velocity and magnetic fields;

\item
A large-scale toroidal magnetic field is generated by the turbulent cross-helicity effect coupled with the large-scale vortical or rotational motions;

\item
A large-scale poloidal magnetic field is generated from the toroidal magnetic field through the $\alpha$ or helicity effect reflecting the helical property of turbulence coupled with the large-scale magnetic field;

\item
The poloidal magnetic field generated by the $\alpha$ or helicity effect is coupled with the large-scale vortical or rotational motions, then contributes to generate the turbulent cross helicity whose sign is opposite to the original turbulent cross helicity;

\item
Staring with the turbulent cross helicity of a reversed sign, a toroidal magnetic field whose polarity is reversed as compared to the original one is induced. Then the second half of a cycle of the oscillatory behavior of the magnetic field follows. 
 
\end{itemize}

	In this scenario, the oscillation of the turbulent cross helicity is considered to play an essential role in the magnetic polarity reversal. It is clear that the sign of the cross helicity changes as the magnetic field changes its polarity:
\begin{equation}
	\left\langle {{\bf{u}} \cdot {\bf{b}}} \right\rangle
	\longmapsto \left\langle {{\bf{u}} \cdot -{\bf{b}}} \right\rangle
	= - \left\langle {{\bf{u}} \cdot {\bf{b}}} \right\rangle
	\label{eq:ub_symmetry}
\end{equation}
with ${\bf{b}} \longmapsto - {\bf{b}}$. Then important point to examine is the phase relationship between the turbulent cross helicity and the mean magnetic field: which of the turbulent cross helicity and the mean magnetic field changes its sign in advance to the other. As for this phase relationship, the present results of the eigenvalue analysis show that the sign change of the turbulent cross helicity should proceed to the counterpart of the mean magnetic field. In this sense, the cross-helicity oscillation is not the result of the magnetic polarity reversal, but the cause of the latter. 

	In the present model we adopt the cross-helicity effect coupled with the mean vorticity [the $P^2$-related terms in Eq.~(\ref{eq:final_tor_eq}) originated from the $f_1$- and $f_2$-related terms in Eq.~(\ref{eq:on-off_tor_eq})] but neglect the differential-rotation or $\Omega$ effect [the $f_4$-related term in Eq.~(\ref{eq:on-off_tor_eq})] as the toroidal magnetic-field generation mechanism. This treatment does not deny the importance of the $\Omega$ effect. What we showed here is that some basic features of the solar-activity cycle can be reproduced by the cross-helicity effect even without resorting to the $\Omega$ effect. 
	
	On the other hand, the cross-helicity effect presented in this work seems to be similar to the $\Omega$ effect since both arise from the mean-velocity inhomogeneity [compare the second terms of Eqs.~(\ref{eq:alpha-Omega_tor_eq}) and (\ref{eq:original_tor_eq})].  The cross-helicity effect depends on the presence of the turbulence and of the cross-correlation between the velocity and magnetic-field fluctuations whereas the $\Omega$ effect does not need turbulence itself. Examination of the relative importance of the cross-helicity effect to the $\Omega$ one in a realistic stellar situation, including the solar case, would provide a very interesting subject of study. 
	
	The relative importance of the cross-helicity effect to the differential rotation effect is estimated by
\begin{equation}
	\Gamma = \frac{|\nabla \times (\gamma \mbox{\boldmath$\Omega$})|}{|\nabla \times ({\bf{U}} \times {\bf{B}})|}.
	\label{eq:Gamma_ratio_def}
\end{equation}
From the second terms of Eqs.~(\ref{eq:original_tor_eq}) and (\ref{eq:alpha-Omega_tor_eq}), $\Gamma$ can be expressed as
\begin{equation}
	\Gamma = \frac{\gamma}{\displaystyle k_u \frac{\partial A}{\partial x}}
	\sim \frac{\langle {{\bf{u}}' \cdot {\bf{b}}'} \rangle}{D \left( {\displaystyle{\frac{\partial U}{\partial r}}} \right) B^r} \frac{\tau_{\rm{turb}}}{\tau_{\rm{mean}}}
\sim \frac{\langle {{\bf{u}}' \cdot {\bf{b}}'} \rangle}{\delta U B^r} Ro^{-1},
	\label{eq:Gamma_ratio_est}
\end{equation}
where $D$ is the radial length scale (dimension of convection zone), $\delta U$ is the toroidal velocity difference with respect to the radial direction, $Ro$ is the Rossby number based on differential velocity (ratio of the mean and turbulence timescales, $\tau_{\rm{mean}}$ and $\tau_{\rm{turb}}$) defined by
\begin{equation}
	Ro = \frac{\tau_{\rm{mean}}}{\tau_{\rm{turb}}}
	= \frac{D/\delta U}{K/\varepsilon}
	\label{eq:Rossby_num_def}
\end{equation}
[$K$: turbulent energy defined in Eq.~(\ref{eq:beta_model}), $\varepsilon$: its dissipation rate]. We see from Eq.~(\ref{eq:Gamma_ratio_est}) that the $\Gamma$ ratio can not be negligible if the turbulent cross helicity associated with the mean toroidal fields is comparable with the poloidal magnetic field $B^r$ multiplied by the differential velocity $\delta U$. Of course, in a nearly solid rotation case ($\delta U \simeq 0$), $\Gamma$ becomes very large.

	An estimate of $\Gamma$ using a direct numerical simulation (DNS) of the solar-like convective shell suggests that $\Gamma \lesssim 0.2$ at the bottom of convection zone ($r/r_{\rm{sun}} \le 0.75$). The $\Gamma$ value monotonically increases as the radial position increases, and reaches $\Gamma \gtrsim 2$ near the surface ($r/r_{\rm{sun}} \ge 0.96$). This increase is mainly caused by the increase of the mean vorticity near the surface region. This suggests the cross-helicity effect is important near the surface of the solar convection zone \citep{mie2016}. 
	
	As we saw in \S~\ref{sec:cr_phys}, the physical origin of the cross-helicity effect consists of the vortical and/or rotational motion and the cross-correlation between the velocity and magnetic-field fluctuations. This suggests that the cross-helicity effect arises even in a solid-rotation case while the $\Omega$ effect requires a preferred differential rotation. This is a substantial difference between the two effects. It must be very interesting to apply the present cross-helicity dynamo to an astrophysical body that rigidly rotates in a highly turbulent state, such as a red dwarf (a small and relatively cool star on the main sequence).

\acknowledgments

	On 4 December 2013, one of the authors, Dieter Schmitt, tragically died  after his heart surgery. This was a terrible shock, and we, and many others, have felt an overwhelming loss. May his soul rest in peace at Eddigehausen beloved of him. The rest of authors would like to cordially dedicate this paper to the memory of Dieter. The authors are indebted to Manfred Sch\"{u}ssler, Arnab Rai Choudhuri, and Kirill Kuzanyan for basic references in this topic. Parts of this work were performed during the periods one of the authors (NY) stayed at the Institut f\"{u}r Astrophysik (IAP), Georg-August-Universit\"{a}t G\"{o}ttingen in 2013, the Kiepenheuer-Institut f\"{u}r Sonnenphysik (KIS) and Max-Planck Institut f\"{u}r Sonnensystemforschung (MPS) in 2014, and the Nordic Institute for Theoretical Physics (NORDITA) program on Magnetic Reconnection in Plasmas in 2015. This work was also supported by the Japan Society for the Promotion of Science (JSPS) Grants-in-Aid for Scientific Research (No. 24540228).

\appendix
\section{Detailed expressions for the transport coefficients\label{sec:a1}}
	It is worth while to note that Eq.~(\ref{eq:crss_hel_effect_cal_iso}) is based on the most simplified assumption on the statistical property of turbulence, Eq.~(\ref{eq:crss_corl_iso}). In realistic turbulence, we should adopt more elaborated closure applicable to the inhomogeneous anisotropic magnetohydrodynamic (MHD) turbulence. Results obtained from one of such elaborated closure schemes are seen in \citet{yos1990} and \citet{yok2013a}. These results were obtained without resorting to the quasi-linear or so-called first-order smoothing approximation (FOSA). On the contrary, these results are based on a theory that is appropriate for the fully-nonlinear turbulence with very high kinetic and magnetic Reynolds numbers \citep{kra1959,yos1984,yos1990,yok2013a}. According to the results of such analyses, the primary part of the EMF is expressed by Eq.~(\ref{eq:emf_model_full}), but with much more elaborated expressions for the turbulent transport coefficients such as
\begin{subequations}\label{eq:abc_exp_wave}
\begin{eqnarray}
	\beta &=& \int d{\bf{k}} \int_{-\infty}^t\!\!\!\!\! d\tau'
		G(k,{\bf{x}};\tau,\tau',t)
	\nonumber\\ 
		&&\hspace{-10pt}\left[{
		Q_{uu}(k,{\bf{x}};\tau,\tau',t) + Q_{bb}(k,{\bf{x}};\tau,\tau',t)
		}\right],
	\label{eq:beta_exp_wave}
\end{eqnarray}
\begin{eqnarray}
	\alpha &=& \int d{\bf{k}} \int_{-\infty}^t\!\!\!\!\! d\tau'
		G(k,{\bf{x}};\tau,\tau',t)
	\nonumber\\
		&&\hspace{-10pt}\left[{
		-H_{uu}(k,{\bf{x}};\tau,\tau',t) + H_{bb}(k,{\bf{x}};\tau,\tau',t)
		}\right],
	\label{eq:alpha_exp_wave}
\end{eqnarray}
\begin{eqnarray}
	\gamma &=& \int d{\bf{k}} \int_{-\infty}^t\!\!\!\!\! d\tau'
		G(k,{\bf{x}};\tau,\tau',t)
	\nonumber\\ 
		&&\hspace{-10pt}\left[{
		Q_{ub}(k,{\bf{x}};\tau,\tau',t) + Q_{bu}(k,{\bf{x}};\tau,\tau',t)
		}\right],
	\label{eq:gamma_exp_wave}
\end{eqnarray}
\end{subequations}
where $G$, $Q_{uu}$, $Q_{bb}$, $H_{uu}$, $H_{bb}$, $Q_{ub}$, etc.\ denote the propagators (response or Green's functions and spectral functions) of lowest-order turbulent fields in the wave-number space. The green's function $G$ tells the weight of the past: how much the past affects the present state of turbulence. For example, if we can treat the time and spectral integrals independently, the time integral of the Green's function just gives a timescale of turbulence:
\begin{equation}
	\tau = \int_{-\infty}^t \!\!\!\!\! dt' G({\bf{x}};t,t')
	\label{eq:G_timescale}
\end{equation}
with the spectral integral of the spectral function leading to the turbulent energy density:
\begin{equation}
	K = \int d{\bf{k}} \left[ {
		Q_{uu}(k,{\bf{x}};t) + Q_{bb}(k,{\bf{x}};t)
	} \right].
	\label{eq:spect_K_exp}
\end{equation}
In this specific case, we obtain the simplest expression corresponding to Eq.~(\ref{eq:beta_model}). In this sense, expressions [Eq.~(\ref{eq:abc_exp_wave})] are the natural generalization of the simplest model expressions of Eq.~(\ref{eq:abc_model}).






\begin{thebibliography}{}

\bibitem[Brandenburg \& Subramanian(2005)] {bra2005}
Brandenburg, A. \& Subramanian, K. 
``Astrophysical magnetic fields and nonlinear dynamo theory,''
Phys.\ Reports, 417, 1-209 (2005).

\bibitem[Charbonneau(2010)]{cha2010}
Charbonneau, P. 
``Dynamo models of the solar cycle,'' 
Living Rev.\ Solar Phys., 7, 3-1-91 (2010).

\bibitem[Charbonneau(2014)]{cha2014}
Charbonneau, P. 
``Solar dynamo theory,''
Annu.\ Rev.\ Astron.\ Astrophys., 52, 251-290 (2014).

\bibitem[Parker(1955)]{par1955}
Parker, E. N. 
``Hydromagnetic dynamo models,''
Astrophys.\ J., 122, 293-314 (1955).

\bibitem[Moffatt(1978)]{mof1978}
Moffatt, K. 
Magnetic Field Generation in Electrically Conducting Fluids, 
Cambridge: Cambridge University Press

\bibitem[Parker(1979)]{par1979}
Parker, E. N. 
Cosmical Magnetic Fields, 
Oxford: Oxford University Press

\bibitem[Krause \& R\"{a}dler(1980)]{kra1980}
Krause, F. \& R\"{a}dler, K.-H. 
Mean-field magnetohydrodynamics and dynamo theory, 
(Oxford: Pergamon Press 1980).

\bibitem[Parker(1993)]{par1993}
Parker, E. N. 
``A solar dynamo surface wave at the interface between convection and nonuniform rotation,''
Astrophys.\ J., 408, 707-719 (1993).

\bibitem[Tobias(1996)]{tob1996}
Tobias, S. M. 
``Diffusivity quenching as a mechanism for Parker's surface dynamo,'' 
Astrophys.\ J., 467, 870-880 (1996).

\bibitem[Wang, Sheeley \& Nash(1991)]{wan1991}
Wang,Y. -M., Sheeley, N. R., \& Nash, A. G. 
``A new solar cycle model including meridional circulation,''
Astrophys.\ J., 383, 431-442 (1991).

\bibitem[Chodhuri, Sch\"{u}ssler \& Dikpati(1995)]{cho1995}
Choudhuri, A. R., Sch\"{u}ssler, M., \& Dikpati, M. 
``The solar dynamo with meridional circulation,''
Astron.\ Astrophys., 303, L29-L32 (1995).

\bibitem[Durney(1995)]{dur1995}
Durney, B. R. 
``On a Babcock-Leighton dynamo model with a deep-seated generating layer for the toroidal magnetic field,''
Solar Phys., 160, 213-235 (1995).

\bibitem[Choudhuri(2011)]{cho2011}
Choudhuri, A. R. 
``The origin of the solar magnetic cycle,''
Pramana, J. Phys., 77, 77-96 (2011).

\bibitem[Brandenburg et al.(1989)]{bra1989}
Brandenburg, A., Krause, F., Meinel, R., Moss, D. \& Tuominen, I. 
``The stability of nonlinear dynamos and the limited role of kinematic growth rates,''
Astron.\ Astrophys., 213, 411-422 (1989).

\bibitem[Tobias(1997)]{tob1997}
Tobias, S. M. 
``The solar cycle: parity interactions and amplitude modulation,'' 
Astron.\ Astrophys., 322, 1007-1017 (1997).

\bibitem[Kleeorin et al.(2000)]{kle2000}
Kleeorin, N., Moss, D., Rogachevskii, I., \& Sokoloff, D. 
``Helicity balance and steady-state strength of the dynamo generated galactic magnetic field,''
Astron.\ Astrophys., 361, L5-L8 (2000).

\bibitem[Kleeorin et al.(2006)]{kle2006}
Kleeorin, N., Moss, D., Rogachevskii, I., \& Sokoloff, D. 
``The nonlinear galactic dynamo and magnetic helicity transport,''
Astron.\ Nachr., 327, 473-474 (2006).
 
\bibitem[Pipin(1999)]{pip1999}
Pipin, V. V. 
``The Gleissberg cycle by a nonlinear $\alpha$ $\Lambda$ dynamo,''
Astron.\ Astrophys., 346, 295-302 (1999).

\bibitem[Pipin(2004)]{pip2004}
Pipin, V. V. 
``Variations in the Solar Luminosity, Radius, and Quadrupole Moment as Effects of a Large-Scale Dynamo in the Solar Convection Zone,''
Astronomy Reports, 48, 418-432 (2004).

\bibitem[Pipin \& Kosovichev(2011)]{pip2011a}
Pipin, V. V. \& Kosovichev, A. G. 
``The asymmetry of sunspot cycles and Waldmeier relations as a result of nonlinear surface-shear shaped dynamo,''
Astrophys.\ J., 741, 1-9 (2011).
 
\bibitem[Arlt \& R\"{u}diger(1999)]{arl1999}
Arlt, R. \& R\"{u}diger G. 
``Accretion-disk dynamo models with dynamo-induced alpha effect,''
Astron.\ Astrophys., 349, 334-338 (1999).

\bibitem[Weiss \& Tobias(2016)]{wei2016}
Weiss, N. O. \& Tobias, S. M. 
``Supermodulation of the Sun's magnetic activity: the effects of symmetry changes,'
Mon.\ Not.\ Roy.\ Astron.\ Soc., 456, 2654-2661 (2016).
 
\bibitem[Pouquet, Frisch \& L\'{e}orat(1976)]{pou1976}
Pouquet, A., Frisch, U., \& L\'{e}orat, J. 
``Strong MHD helical turbulence and nonlinear dynamo effect,''
J. Fluid Mech., 77, 321-354 (1976).

\bibitem[Yoshizawa(1990)]{yos1990}
Yoshizawa, A. 
``Self-consistent turbulent dynamo modeling of reversed field pinches and planetary magnetic fields,''
Phys.\ Fluids B, 2, 1589-1600 (1990).

\bibitem[Yokoi(2013)]{yok2013a}
Yokoi, N. 
``Cross helicity and related dynamo,''
Geophys.\ Astrophys.\ Fluid Dyn., 107, 114-184 (2013).

\bibitem[Steenbeck \& Krause(1969)]{ste1969}
Steenbeck, M. \& Krause, F. 
``Zur Dynamotheorie stellarer und planetarer MagnetfelderI. Berechnung sonnenn\"{a}hnlicher Wechselfeldgeneratoren,''
Astron.\ Nachr.\ 291, 49-84 (1969).

\bibitem[Yoshimura(1975a)]{ysmr1975a}
Yoshimura, H. 
``Solar-cycle dynamo wave propagation,''
Astrophys.\ J., 201, 740-748 (1975).

\bibitem[Yoshimura(1975b)]{ysmr1975b}
Yoshimura, H. 
``A model of the solar cycle driven by the dynamo action of the global convection in the solar convection zone,''
Astrophys.\ J.s, 29, 467-494 (1975).

\bibitem[Parker(1975)]{par1975}
Parker, E. N. 
``The generation of magnetic fields in astrophysical bodies. X - Magnetic buoyancy and the solar dynamo,''
Astrophys.\ J., 198, 205-209 (1975).

\bibitem[Spiegel \& Weiss(1980)]{spi1980}
Spiegel, E. A. \& Weiss, N. O. 
``Magnetic activity and variations in solar luminosity,''
Nature, 287, 616- 617 (1980).

\bibitem[van Ballegooijen(1982)]{bal1982}
van Ballegooijen, A. A. 
``The overshoot layer at the base of the solar convective zone and the problem of magnetic flux storage,''
Astron.\ Astrophys., 113, 99-112 (1982).

\bibitem[Dikpati \& Gilman(2006)]{dik2006}
Dikpati M. \& Gilman, P. A. 
``Simulating and predicting solar cycles using a flux-transport dynamo,''
Astrophys.\ J., 649, 498-514 (2006). 

\bibitem[D'silva \& Choudhuri(1993)]{dsi1993}
D'silva, S. \& Choudhuri A. R. 
``A theoretical model for tilts of bipolar magnetic regions,''
Astron.\ Astrophys., 272, 621-633 (1993).

\bibitem[Fan, Fisher \& DeLuca(1993)]{fan1993}
Fan, Y., Fisher, G. H., \& DeLuca, E. E. 
``The origin of morphological asymmetries in bipolar active regions,''
Astrophys.\ J., 405, 390-401 (1993).

\bibitem[Caligari, Moreno-Insertis \& Sch\"{u}ssler(1995)]{cal1995}
Caligari, P., Moreno-Insertis, F., \& Sch\"{u}ssler, M. 
``Emerging flux tubes in the solar convection zone. I. Asymmetry, tilt, and emergence lattitude,''
Astrophys.\ J., 441, 886-902 (1995).

\bibitem[Stenflo \& Kosovichev(2012)]{ste2012a}
Stenflo, J. \& Kosovichev, A. G. 
``Bipolar magnetic regions on the Sun: Global analysis of the SOHO/MDI data set,''
Astrophys.\ J., 745, 129-140 (2012).

\bibitem[Stenflo(2012)]{ste2012b}
Stenflo, J. 
``Scaling laws for magnetic fields on the quiet Sun,'' 
Astron.\ Astrophys., 541, A17-1-12 (2012).

\bibitem[Tlatov et al.(2013)]{tla2013}
Tlatov, A., Illarinov. E, Sokoloff, D. \& Pipin, V. 
``A new dynamo pattern revealed by the tilt angle of bipolar sunspot groups,''
Mon.\ Not.\ Roy.\ Astron.\ Soc., 432, 2975-2984 (2013).

\bibitem[McClintock \& Norton(2016)]{mcc2016}
McClintock, B. H. \& Norton, A. A. 
``Tilt Angle and footpoint separation of small and large bipolar sunspot regions observed with HMI,''
Astrophys.\ J., 818, 7-1-7 (2016).

\bibitem[Benevolenskaya et al.(1999)]{ben1999}
Benevolenskaya, E. E., Hoeksema, J. T., Kosovichev, A. G., \& Scherrer, P. H. 
``The interaction of new and old magnetic fluxes at the beginning of solar cycle 23,''
Astrophys.\ J.l, 517, L163-L166 (1999).

\bibitem[Dasi-Espuig et al.(2010)]{das2010}
Dasi-Espuig, M., Solanki, S. K., Krivova, N. A., Cameron R., \& Pe\~{n}uela, T. 
``Sunspot group tilt angles and the strength of the solar cycle,''
Astron.\ Astrophys., 518, A7-1-10 (2010).

\bibitem[Bonanno, Elstner \& Belvedere(2006)]{bon2006}
Bonanno, A., Elstner, D., \& Belvedere, G. 
``Advection-dominated solar dynamo model with two-cell meridional flow and a positive $\alpha$-effect in the tachocline,''
Astron.\ Nachr., 327, 680-685 (2006).

\bibitem[Jouve \& Brun(2007)]{jou2007}
Jouve, L. \& Brun, A. S. 
``On the role of meridional flows in flux transport dynamo models,''
Astron.\ Astrophys., 474, 239-250 (2007).

\bibitem[Zhao et al.(2013)]{zha2013}
Zhao, J., Bogart, R. S., Kosovichev, A. G., Duvall, Jr., T. L., \& Hartlep, T. 
``Detection of equatorward meridional flow and evidence of double-cell meridional circulation inside the sun,''
Astrophys.\ J.l, 774, L29-L34 (2013).

\bibitem[Schad, Timmer \& Roth(2013)]{sch2013}
Schad, A., Timmer, J., \& Roth, M. 
``Global helioseismic evidence for a deeply penetrating solar meridional flow consisting of multiple flow cells,''
Astrophys.\ J.l, 778, L38-L44 (2013).

\bibitem[Rajaguru \& Antia(2015)]{raj2015}
Rajaguru, S. P. \& Antia, H. M. 
``Meridional circulation in the solar convection zone: Time-distance helioseismic inferences from four years of HMI/SDO observations,''
Astrophys.\ J., 813, 114-121 (2015).

\bibitem[Hazra, Karak \& Choudhuri(2014)]{haz2014}
Hazra, G., Karak, B. B., \& Choudhuri, A. R. 
``Is a deep one-cell meridional circulation essential for the flux transport solar dynamo?''
Astrophys.\ J., 782, 93-104 (2014).
 
 \bibitem[Pipin \& Kosovichev(2013)]{pip2013}
Pipin, V. V. \& Kosovichev, A. G. 
``The mean-field solar dynamo with a double cell meridional circulation pattern,'' 
Astrophys.\ J., 776, 36-44 (2013)

 \bibitem[Pipin \& Kosovichev(2014)]{pip2014a}
Pipin, V. V. \& Kosovichev, A. G. 
``Effects of anisotropies in turbulent magnetic diffusion in mean-field solar dynamo models,'' 
Astrophys.\ J., 785, 49-60 (2014)

\bibitem[Racine et al.(2011)]{rac2011}
Racine, \'{E}., Charbonneau, P., Ghizaru, M., Bouchat, A., \& Smolarkiewicz, P. K. 
``On the mode of dynamo action in a global large-eddy simulation of solar convection,''
Astrophys.\ J., 735, 46-67 (2011).

\bibitem[Warnecke et al.(2014)]{war2014}
Warnecke, J., K\"{a}pyl\"{a}, P. J., K\"{a}pyl\"{a}, M. J., \& Brandenburg, A. 
``On the cause of solar-like equatorward migration in global convective dynamo simulations,''
Astrophys.\ J.l, 796, L12-L17 (2014).

\bibitem[Choudhuri, Chatterjee \& Jiang(2007)]{cho2007}
Choudhuri, A. R.,  Chatterjee, P.,  \& Jiang, J. 
``Predicting Solar Cycle 24 With a Solar Dynamo Model,''
Phys.\ Rev.\ Lett., 98, 131103-1-4 (2007).

\bibitem[Pipin et al.(2014)]{pip2014b}
Pipin, V. V., Moss, D., Sokoloff, D., \& Hoeksema, T. 
``Reversals of the solar magnetic dipole in the light of observational data and simple dynamo models,''
Astron.\ Astrophys., 567, A90-1-9 (2014).

\bibitem[Schlichenmaier \& Stix(1995)]{sch1995}
Schlichenmaier, R. \& Stix, M. 
``The phase of the radial mean field in the solar dynamo,''
Astron.\ Astrophys., 302, 264-270 (1995).

\bibitem[Kleeorin, Rogachevskii \& Ruzmaikin(1995)]{kle1995}
Kleeorin, N., Rogachevskii, I., \& Ruzmaikin, A. 
``Magnitude of the dynamo-generated magnetic field in solar-type convective zones,''
Astron.\ Astrophys., 297, 159-167 (1995).

\bibitem[Biskamp(1993)]{bis1993}
Biskamp, D. 
{\it Nonlinear Magnetohydrodynamics}, 
(Cambridge: Cambridge University Press, 1993)

\bibitem[Yokoi et al.(2008)]{yok2008}
Yokoi, N., Rubinstein, R., Yoshizawa, A., \& Hamba, F. 
``A turbulence model for magnetohydrodynamic plasmas,''
J. Turbulence, 9, N37-1-25 (2008).

\bibitem[Yokoi(2011)]{yok2011b}
Yokoi, N. 
``Modeling the turbulent cross-helicity evolution: production, dissipation, and transport rates,''
J. Turbulence, 12, N27-1-33 (2011).

\bibitem[Yokoi \& Hoshino(2011)]{yok2011c}
Yokoi, N. \& Hoshino, M. 
``Flow--turbulence interaction in magnetic reconnection,''
Phys.\ Plasmas, 18, 11208-1-14 (2011).

 \bibitem[Higashimori, Yokoi \& Hoshino(2013)]{hig2013}
 Higashimori, K., Yokoi, N., \& Hoshino, M. 
 ``Explosive turbulent magnetic reconnection,''
 Phys.\ Rev.\ Lett., 110, 255001-1-5 (2013).
 
\bibitem[Yokoi, Higashimori \& Hoshino(2013)]{yok2013b}
Yokoi, N., Higashimori, K., \& Hoshino, M. 
``Transport enhancement and suppression in turbulent magnetic reconnection: A self-consistent turbulence model,''
Phys.\ Plasmas, 20, 122310-1-17 (2013)

 \bibitem[Widmer, B\"{u}chner \& Yokoi(2016)]{wid2016}
 Widmer, F., B\"{u}chner, J., \& Yokoi, N. 
 ``Subgrid-scale description of turbulent magnetic reconnection in magnetohydrodynamics,''
 submitted to Phys.\ Plsamas, arXiv:1511.04347

\bibitem[Kleeorin et al.(2003)]{kle2003}
Kleeorin, N., Kuzanyan, K., Moss, D., Rogachevskii, I., Sokoloff, D., \& Zhang, H. 
``Magnetic helicity evolution during the solar activity cycle: Observations and dynamo theory,''
Astron.\ Astrophys., 409, 1097-1105 (2003).

\bibitem[Kuzanyan, Pipin \& Zhang(2007)]{kuz2007}
Kuzanyan, K. M., Pipin, V. V., \& Zhang,  H. 
``Probing current and cross-helicity in the solar atmosphere: A challenge for theory,''
Adv. Space Res., 39, 1694-1699 (2007).

\bibitem[Pipin et al.(2011)]{pip2011b}
Pipin, V. V., Kuzanyan, K. M., Zhang, H., \& Kosovichev, A. G. 
``Turbulent cross-helicity in the mean-field solar dynamo problem,''
Astrophys.\ J., 743, 160-1-12 (2011).

\bibitem[Zhao, Wang \& Zhang(2011)]{zha2011}
Zhao, M. Y., Wang, X. F., \& Zhang, H. Q. 
``The correlation between the magnetic and velocity fields on the full solar disk,''
Solar Phys., 270, 23-33 (2011).

\bibitem[Yoshizawa, Kato \& Yokoi(2000)]{yos2000}
Yoshizawa, A., Kato, H., \& Yokoi, N. 
``Mean field theory interpretation of solar polarity reversal,''
Astrophys.\ J., 537, 1039-1053 (2000).

\bibitem[Yoshizawa(1984)]{yos1984}
Yoshizawa, A. 
``Statistical analysis of the deviation of the Reynolds stress from its eddy-viscosity representation,''
Phys.\ Fluids, 27, 1377-1387 (1984).

\bibitem[Yokoi \& Yoshizawa(1993)]{yok1993}
Yokoi, N. \& Yoshizawa, A. 
``Statistical analysis of the effects of helicity in inhomogeneous turbulence,''
Phys.\ Fluids A, 5,  464-477 (1993).

\bibitem[Yokoi \& Brandenburg(2016)]{yok2016}
Yokoi, N. \& Brandenburg, A. 
``Large-scale flow generation by inhomogeneous helicity,''
Phys.\ Rev.\ E, 93, 033125-1-14 (2016).

\bibitem[Yokoi \& Balarac(2011)]{yok2011a}
Yokoi, N. \& Balarac, G. 
``Cross-helicity effects and turbulent transport in magnetohydrodynamic flow,''
J. Physics: Conf.\ Ser., 318, 072039-1-10 (2011).

\bibitem[Miesch(2016)]{mie2016}
Miesch, M. 
Private communication (2016).

\bibitem[Kraichnan(1959)]{kra1959}
Kraichnan, R. H. 
``The structure of isotropic turbulence at very high Reynolds numbers,''
J. Fluid Mech., 5, 497-543 (1959).


\end{thebibliography}
\end{document}